\documentclass{aa}
\usepackage{psfig,graphicx,ulem,epsfig}
\usepackage{txfonts}
\usepackage{morefloats}

\begin{document}

\title{The optical properties of galaxies in the Ophiuchus cluster }

\author{
F. Durret \inst{1} \and
K.~Wakamatsu \inst{2} \and
C.~Adami \inst{3} \and
T.~Nagayama  \inst{4} \and 
J.M.~Omega Muleka Mwewa Mwaba \inst{5,6}
}

\institute{
Sorbonne Universit\'e, CNRS, UMR~7095, Institut d'Astrophysique 
de Paris, 98bis Bd Arago, 75014, Paris, France
\and
Faculty of Engineering, Gifu University, 1-1 Yanagido, Gifu 501-1193, Japan
\and
LAM, OAMP, P\^ole de l'Etoile Site Ch\^ateau-Gombert, 38 rue Fr\'ed\'eric 
Joliot--Curie,  13388 Marseille Cedex 13, France
\and
Department of Astrophysics, Kagoshima University, Nagoya, Japan
\and
Universidade Federal de Sergipe, Departamento de F\'isica, Av. Marechal Rondon, S/N, 49000-000 S\~ao Crist\'ov\~ao, SE, Brazil
\and
CAPES Foundation, Ministry of Education of Brazil, Brasilia \^a DF, Zip Code 70.040-020,
Brazil
  }

\date{Accepted . Received ; Draft printed: \today}

\authorrunning{Durret et al.}

\titlerunning{Galaxies in the Ophiuchus cluster}

\abstract
{Ophiuchus is one of the most massive clusters known, but due to its
low Galactic latitude its optical properties remain poorly known. }
{ We investigate the optical properties of Ophiuchus to obtain clues
  on the formation epoch of this cluster, and compare them to those of the
  Coma cluster, which is comparable in mass to Ophiuchus but much
  more disturbed dynamically.}
{Based on a deep image of the Ophiuchus cluster in the $r'$ band
  obtained at the Canada France Hawaii Telescope with the MegaCam
  camera, we have applied an iterative process to subtract the
  contribution of the numerous stars that pollute the image, due to
  the low Galactic latitude of the cluster, and obtained a photometric
  catalogue of 2818 galaxies fully complete at $r'=20.5$~mag and
  still 91\% complete at $r'=21.5$~mag. We use this catalogue to
  derive the cluster Galaxy Luminosity Function (GLF) for the overall
  image and for a region (hereafter the ``rectangle'' region) covering
  exactly the same physical size as the region in which the GLF of the
  Coma cluster was studied by Adami et al. (2007).  We then compute
  density maps based on an adaptive kernel technique, for different
  magnitude limits, and define three circular regions covering 0.08,
  0.08 and 0.06~deg$^2$ respectively centered on the cluster (C),
  northwest (NW) and southeast (SE) of the cluster, in which we
  compute the GLFs.  }
{The GLF fits are much better when a Gaussian is added to the usual
  Schechter function, to account for the excess of very bright
  galaxies. Compared to Coma, Ophiuchus shows a strong excess of
  bright galaxies.}
{The properties of the two nearby very massive clusters Ophiuchus and
  Coma are quite comparable, though they seem embedded in different
  large scale environments. Our interpretation is that Ophiuchus has
  built up long ago, as confirmed by its relaxed state (see paper~I)
  while Coma is still in the process of forming.  }


\maketitle

\section{Introduction}

The Ophiuchus cluster is one of the most massive nearby clusters, at a
redshift of $z=0.0296$. Since its discovery (Wakamatsu \& Malkan 1981,
Johnston et al. 1981), it has been studied in detail in X-rays (see
Werner et al. 2016 and references therein), but due to its low
Galactic latitude ($b=9.3^\circ$) its optical properties are not well
known. In a previous paper (Durret et al. 2015, hereafter paper I), we
have shown that overall the cluster can be considered as relaxed,
composed of a main structure and a single much smaller substructure,
with a total mass of $1.1\ 10^{15}$~M$_\odot$.  We present here our
photometric catalogue of 2818 galaxies in the $r'$ band, and
discuss the galaxy luminosity function (GLF) in the overall image, in
a region (hereafter the ``rectangle'' region) covering exactly the
same physical size as the region in which the GLF of the Coma cluster
was studied by Adami et al. (2007), and in three regions of the
cluster.

In spite of numerous studies, GLFs do not seem to have properties that
can be predicted based on the simple knowledge of the cluster mass or
structure (relaxed or not). GLFs are usually fit by a Schechter
function (Press \& Schechter 1974) but deviations from this function
are often observed, in particular in merging clusters, which often
show an excess of very bright galaxies. At the other end of the GLF,
the faint end slope $\alpha$ gives us informations on the cluster
formation and evolution (see e.g. Martinet et al. 2015). However, we
can note that $\alpha$ seems to vary from one cluster to another with
no obvious dependence on the cluster properties. Deriving the GLF of
Ophiuchus and comparing it to that of other massive nearby clusters
such as Coma, which shows a much higher level of substructuring, could
reveal interesting differences or similarities between two clusters of
comparable masses, one relaxed and the other non-relaxed.

In a preliminary study based on an extensive redshift catalogue
combining their own data with archive data from 6dF and NED, Wakamatsu
et al. (2005) have analysed the distribution of 4717 galaxies with
recession velocities in the range $500<cz<40,000$~km~s$^{-1}$. They
showed that Ophiuchus is sufficiently large and rich to be considered
as a supercluster, which they call the Ophiuchus supercluster.
Wakamatsu et al. (2005) also detected a wall structure 65~Mpc long
between Ophiuchus and a zone from where the Hercules supercluster
extends to the north. They noted that this wall runs close to
north-south and crosses the Great Wall perpendicularly at the Hercules
supercluster. Another wall could be a continuation of the
Ophiuchus-Hercules wall across and beyond the Galactic plane. They
also found a strong deficiency of galaxies with velocities
$cz<4000$~km~s$^{-1}$, thus extending the Local Void beyond the limit
determined by Tully \& Fisher (1987). So Ophiuchus seems to show
particularities at large scales, and this will be investigated in a
companion paper (Wakamatsu et al. in preparation).
 
As in paper~I, we adopt the following quantities: cluster centre
RA=258.1155$^\circ$, DEC=$-23.3698^\circ$ (coordinates of the cD
galaxy), scale of 0.585 kpc/arcsec, distance modulus of 35.54.

\section{The photometric galaxy catalogue}

The study presented here is based on the full exploitation of the
CFHT/MegaCam image that we obtained in the $r'$ band. Magnitudes have
been measured in the AB photometric system. At the redshift of
Ophiuchus, the $1\times 1$ deg$^2$ MegaCam field corresponds to a
region of about $2.1\times 2.1$~Mpc$^2$. This covers part of the
cluster (for which $r_{200}=2.1$~Mpc (see Table~2 in Paper~I).  As
explained in Paper I for our pilot survey in a small area
($10\times 9.5$~arcmin$^2$) centered on the cD galaxy, an automated
galaxy survey did not work properly for the Ophiuchus cluster due to
the high density of foreground stars.  Instead, we made our galaxy
survey based on the eye-inspection of star-subtracted images, which
are created from original ones with a PSF-deblending algorithm.  After
this process, galaxies are fairly easily
found.
To make the survey as uniform as possible for the present large sky
area (about 1~deg$^2$), we set the limiting magnitude about 1.5~mag
brighter than the previous pilot survey and iterated the survey four
times, going to fainter magnitudes step by step.  This work was done
by one of us (K.W.) and his research assistant. Finally we picked 2818
galaxies, including 227 galaxies that were in the pilot survey of
Paper I.

We now produce a full $r'$ band magnitude catalogue for the entire
1~deg$^2$ region covered by MegaCam, including the galaxies from
Paper~I, and containing 2818 galaxies.  Since our observations
  were made in 2010, the $r'$ band filter mounted on MegaCam was a
  first generation filter. Magnitudes in this filter (in the AB
  system) are very close to the SDSS magnitudes: as indicated in the
  MegaCam
  pages\footnote{http://www.cadc-ccda.hia-iha.nrc-cnrc.gc.ca/en/megapipe/docs/filt.html},
  the relation between the MegaCam and SDSS $r$ band magnitudes is:
  $$r_{MegaCam}-r_{SDSS}=-0.024\ (g_{SDSS}-r_{SDSS}).$$  
  For elliptical galaxies at redshift $z=0$ (the galaxy type expected
  to be dominant in Ophiuchus), Fukugita et al. (1995) give
  $g_{SDSS}-r_{SDSS}=0.77$, in which case
  $r_{MegaCam}=r_{SDSS}-0.013$. For other types of galaxies, the
  difference between $r_{MegaCam}$ and $r_{SDSS}$ will be even
  smaller. In view of the error bars on magnitudes (see
  Table~\ref{tab:errmag}) and of the large extinction correction that
  must be applied (see below), we will hereafter neglect the
  difference between $r_{MegaCam}$ and $r_{SDSS}$.  All the galaxies
  were treated as described in Paper~I (iterative star subtraction)
  and subimages were extracted around each galaxy after the stars were
  eliminated.  The varying extinction in the field results in a
  different sensitivity to galaxy intrinsic surface brightness with
  position. This will impact photometric methods based on a
  thresholding in surface brightness for object measurements, such as
  isophotal magnitudes or Kron magnitudes. For this reason, we have
  used SExtractor MAG\_MODEL magnitudes, based on a profile fit,
  rather than the classical MAG\_AUTO.  The PSF was measured with
  PSFEx\footnote{http://www.astromatic.net/}, and a Sersic profile
  convolved with this PSF was fit to each galaxy using SExtractor
  (Bertin \& Arnouts 1996), giving MAG\_MODEL magnitudes.

\begin{figure}
 \begin{center}
\includegraphics[width=\columnwidth, angle=0]{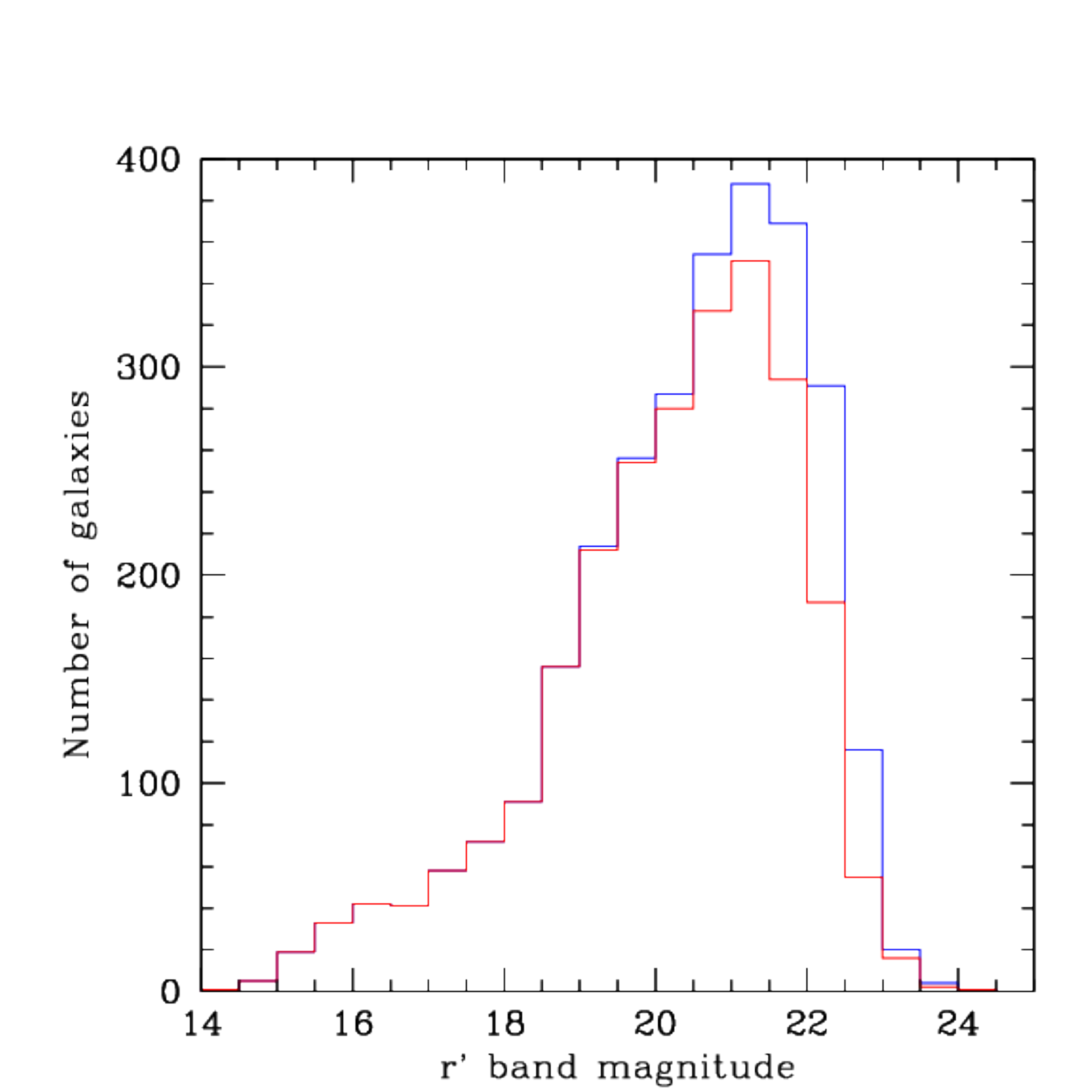}
\caption{Observed magnitude histogram in the $r'$ band for the
  galaxies included in the 1~deg$^2$ region covered by our MegaCam
  image (before applying any extinction correction). The 2499
    galaxies of the preliminary survey are shown in red and the 319
    galaxies of our second survey are shown in blue.  }
\label{fig:histomag}
\end{center}
\end{figure}

\begin{figure}
 \begin{center}
\includegraphics[width=\columnwidth, angle=0]{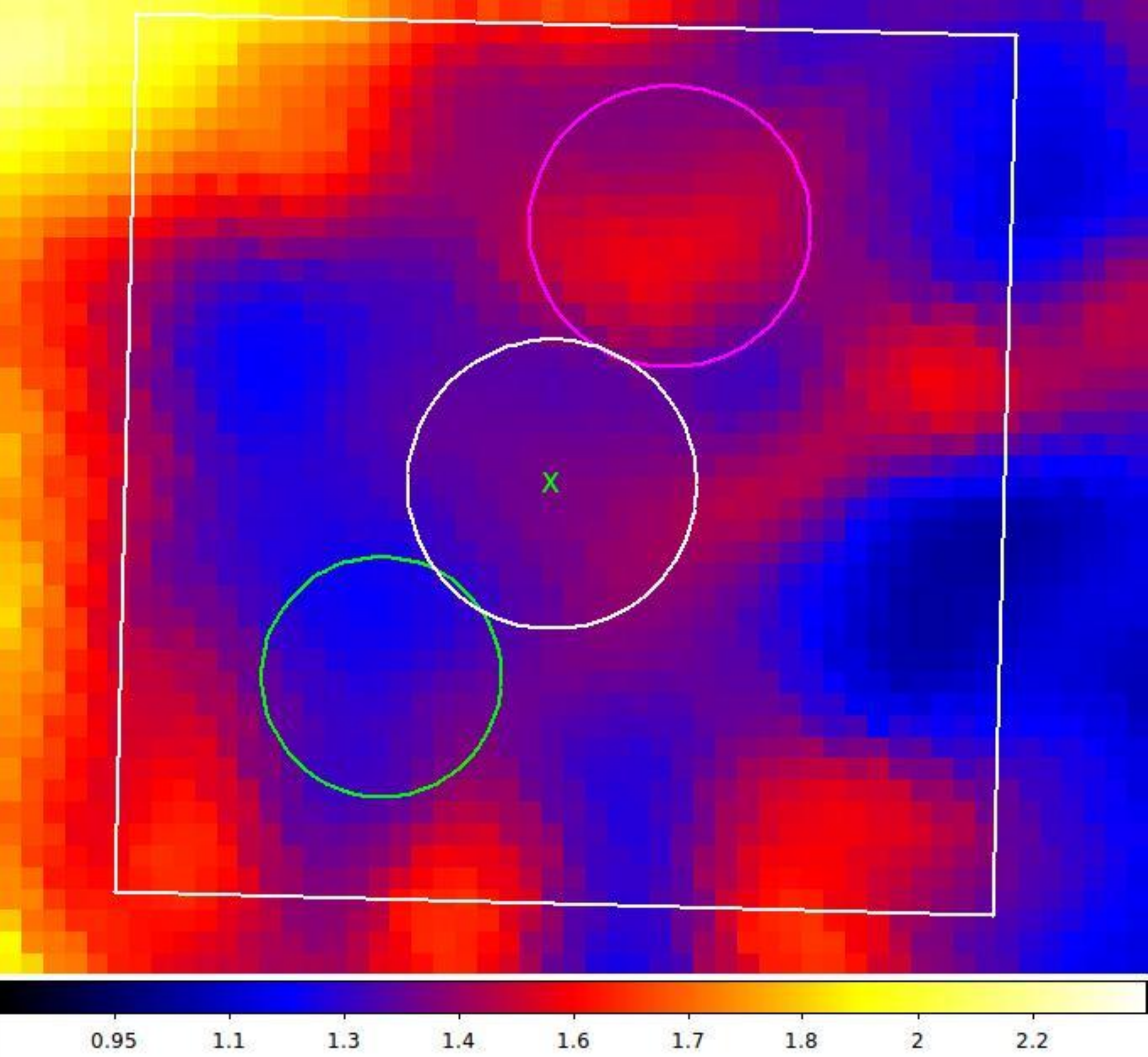}
\caption{Galactic extinction map in the Ophiuchus region. The white
  square shows the MegaCam field, and the three circles are those
  defined in Section~\ref{subsec:densmaps}. Extinction increases from
  blue to red and to yellow. The MegaCam field is oriented with North
  to the top and east to the left, and the extinction map is
  slightly tilted.}
\label{fig:mapexti}
\end{center}
\end{figure}

\begin{figure}
 \begin{center}
\includegraphics[width=\columnwidth, angle=0]{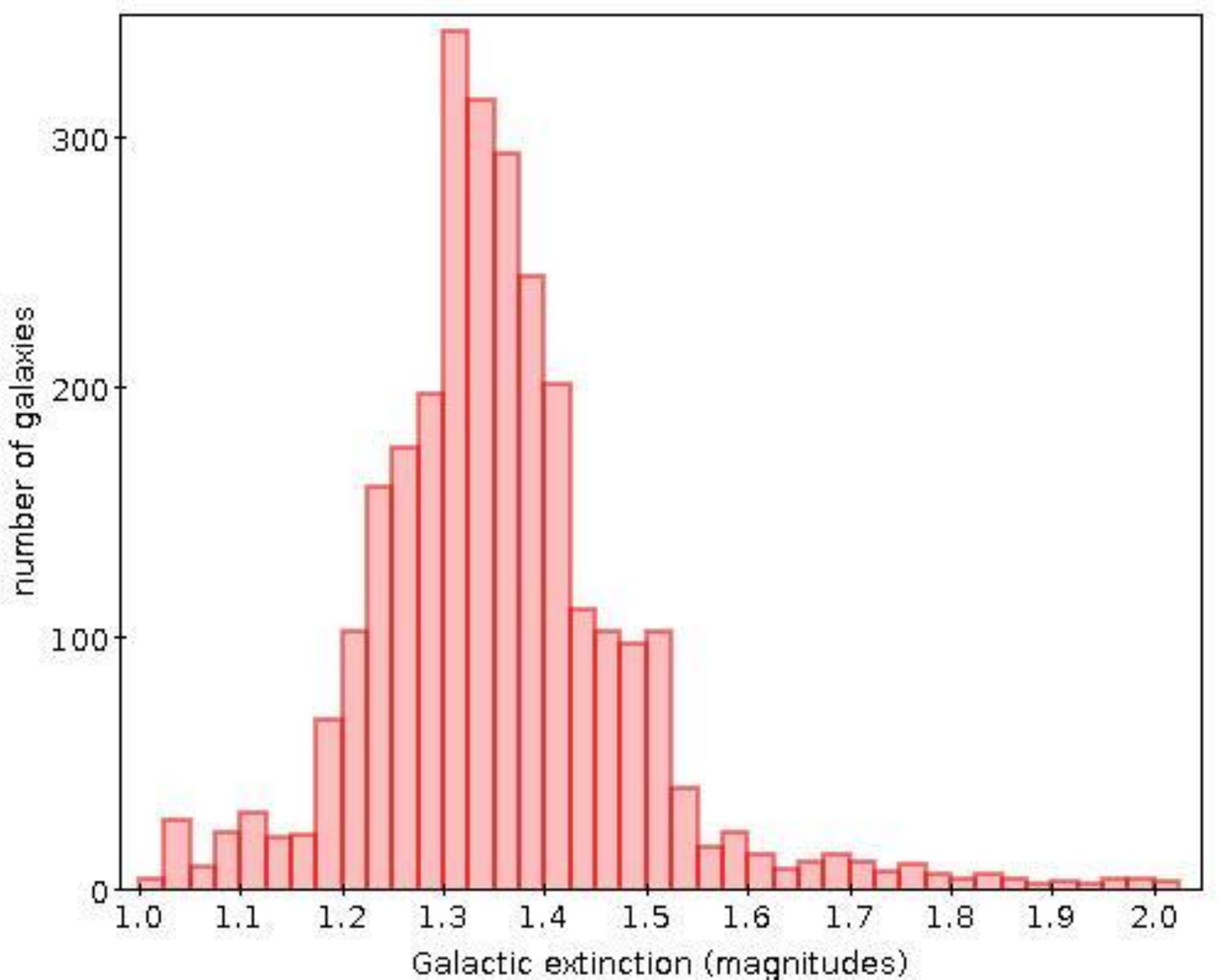}
\caption{Histogram of the Galactic extinction in the $r'$ band for the
  2818 galaxies of our photometric catalogue. }
\label{fig:histoexti}
\end{center}
\end{figure}

To illustrate the depth of the photometric catalogue, we show in
Fig.~\ref{fig:histomag} the corresponding apparent magnitude
histogram.  Nearly half of the galaxies in the three faintest bins in
Fig.~\ref{fig:histomag} are galaxies found in the pilot survey given
in Paper~I, implying that the present galaxy survey in the large area
is shallower than the pilot survey.

To check the completeness of our preliminary galaxy survey containing
2499 galaxies, we marked on the star-subtracted images the detected
galaxies with $20.0 <r'<22.0$ with different symbols in 0.5~mag
intervals, and made the galaxy survey again for the full 1~deg$^2$
area.  We did our best not to miss them during the survey process by
changing the contrast and/or brightness level on the ds9 image
viewer. Out of these newly detected galaxies ($\sim 450$), photometric
measurements were made for 319 galaxies among the brightest.  The
present catalog therefore comprises 2499+319= 2818 objects.

The magnitudes of the 5 brightest galaxies among these newly detected
objects are 19.44, 19.85, 19.90, 20.13, and 20.33 mag respectively.
Also follow 12, 14, 13, and 24 galaxies, in the successive magnitude
ranges from 20.5 to 21.5 with a step of 0.25 mag.  One quarter of the
319 galaxies are brighter than 21.5 mag, while three quarters are
fainter than this magnitude (Fig.~\ref{fig:histomag}).  Based on these
counts, we estimate that the catalog is fully complete down to 20.5
mag, and almost complete (98.5\%) down to 20.75 mag. However, when
the sensitivity variations among the 25 CCD detectors of MegaCam as
well as the existence of faint diffuse nebular emission in this area
of the Galactic plane are taken into account, there may still exist
missed galaxies brighter than 20.5 mag, e.g., diffuse galaxies of low
surface brightness or compact galaxies of small ($\sim 2.5$~arcsec)
angular diameter.
The former may be missed due to irregularities of the sky background,
while the latter can be hidden by the spiders of bright foreground
stars or by multiply blended stars.  We finally estimate that the
present catalog is really complete down to $r'=20.25\pm 0.25$~mag. The
completeness in absolute magnitude is discussed in section 3.1.

We can note that the shortage of galaxies fainter than $M_{r'}=-20$ in
zone C is real (see Sect.~3.4), and is not due to the incompleteness
of our survey in the core region.

At such a low Galactic latitude, Galactic extinction is obviously
large. We first adopted the value of 1.357~mag given by NED for the
SDSS $r$ band.
We then decided to go one step further and to look in more detail at the
Galactic dust extinction map by Schlegel et al. (1998)
with newer estimates from Schlafly and Finkbeiner
(2011)\footnote{http://irsa.ipac.caltech.edu/applications/DUST/}.  The
map contains the values of the colour excess $E(B-V)$ with a pixel
size of 1.5~arcmin.  We converted it to an extinction map $A_r$ in the
$r'$ band, by multiplying it by $R=2.285$ (value taken from Schlafly
\& Finkbeiner 2011, Table~6), i.e., $A_r = R \times E(B-V)$.

The extinction varies quite strongly in the region covered by our
image: between 1.025 and 2.02, as illustrated in
Fig.~\ref{fig:mapexti}. When computing GLFs in Section~\ref{sec:GLF}
we therefore applied either the constant extinction of 1.357~mag or
the individual extinction for each galaxy.

The heavy galactic extinction for the Ophiuchus cluster causes a
  reduction of the apparent angular size of each galaxy, which could
  require further extinction correction. However, this correction
  amounts to 0.15 mag at most, and it depends in a complicated way on
  the adopted color band and morphological type of each galaxy
  (Cameron 1990), so we decided not to apply it.

The final photometric catalogue of 2818 galaxies will be made
available in ViZieR at the
CDS\footnote{http://vizier.u-strasbg.fr/viz-bin/VizieR}.  For each
galaxy, it contains the following information: RA, DEC, measured $r'$
band magnitude with no extinction correction, magnitude corrected for
a constant value of 1.357~mag, magnitude corrected for the individual
extinction of each galaxy, major and minor axes ($a$ and $b$),
position angle (PA) of the major axis, and error on the PA. The
  values of $a$ and $b$ are the angular sizes of the semi-major and
  -minor axes which are computed from the $1\sigma$ values of the
  half-axes given by SExtractor, multiplied by a factor 4.0 to fully
  enclose the galaxies (as advised in the SExtractor manual).  The
  typical errors on the magnitudes, estimated from a subsample of more
  than 40 galaxies measured twice (in adjacent fields), are given in
  Table~\ref{tab:errmag}.

The first ten lines of the catalogue are shown in the Appendix.

\section{The galaxy luminosity function of the Ophiuchus cluster}
\label{sec:GLF}

\subsection{The global galaxy luminosity function}
\label{sec:GLFglobal}

\begin{figure}
 \begin{center}
\includegraphics[width=\columnwidth, angle=0]{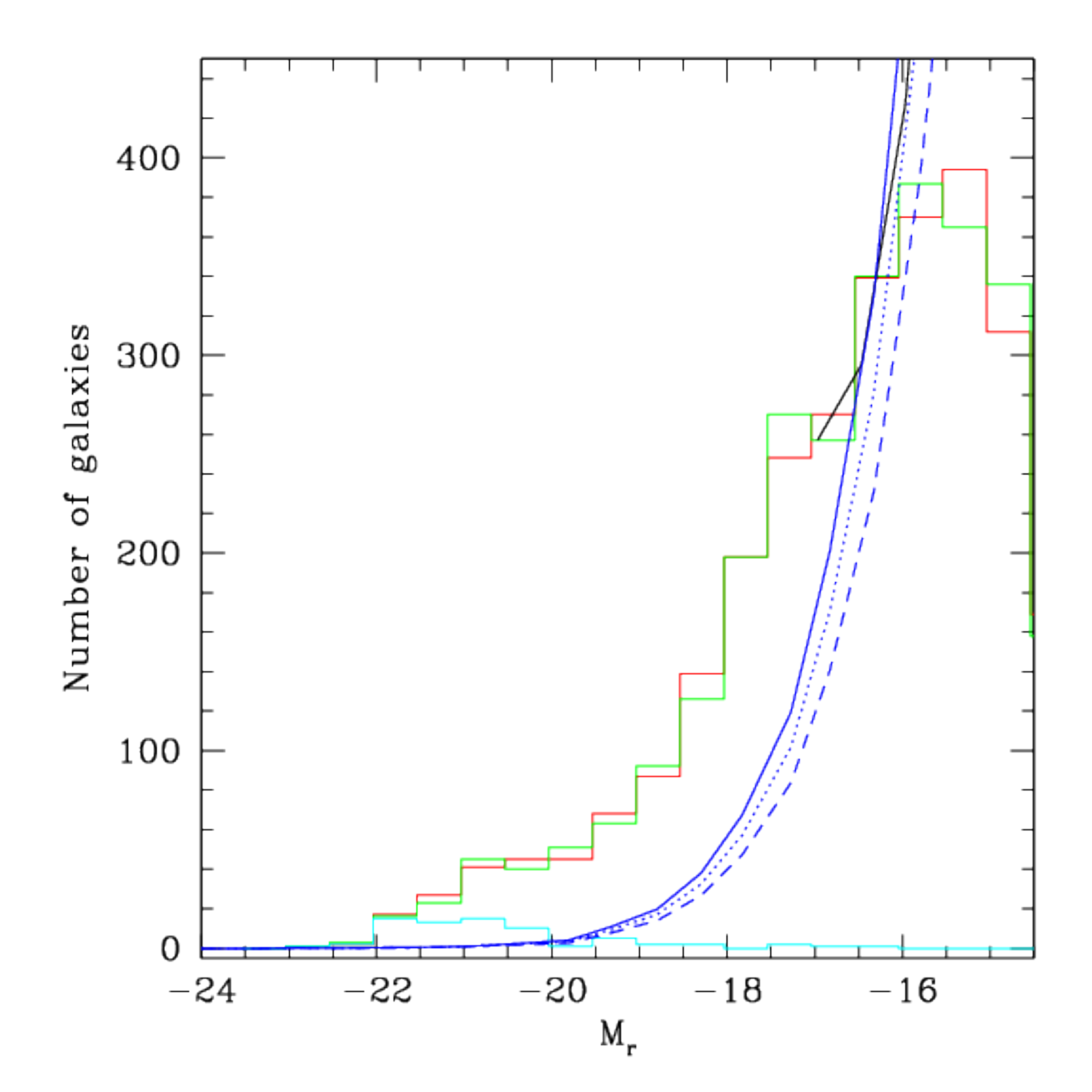}
\caption{Galaxy counts in the $r'$ band converted to absolute
  magnitude (see text). The red histogram shows the 2818 galaxies from
  our catalogue and the cyan histogram those with spectroscopic
  redshifts in the cluster range, both corrected for a constant
  galactic extinction of 1.357 as explained in Sect.~2. The green
  histogram shows the 2818 galaxies from our catalogue corrected for
  their individual galactic extinction. The blue and black lines show
  the field galaxy counts computed by Yasuda et al. (2001), and
  McCracken et al. (2003) respectively. The blue dotted and
    dashed lines show the Yasuda background counts multiplied by 0.85
    and 0.70, respectively (see Sect.~3.1).}
\label{fig:counts}
\end{center}
\end{figure}

\begin{figure}
 \begin{center}
\includegraphics[width=4cm, angle=0]{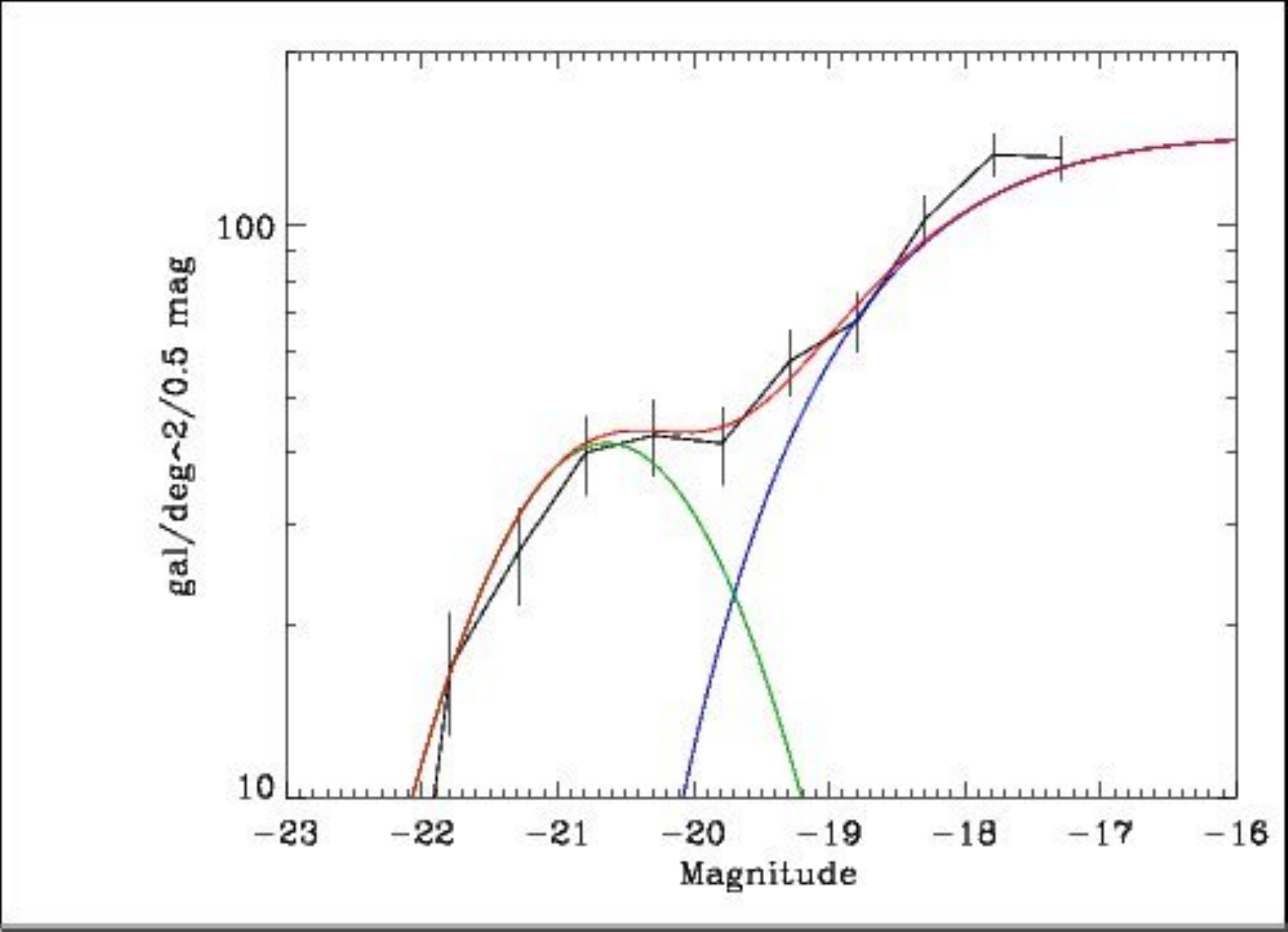}\kern0.2cm%
\includegraphics[width=4cm, angle=0]{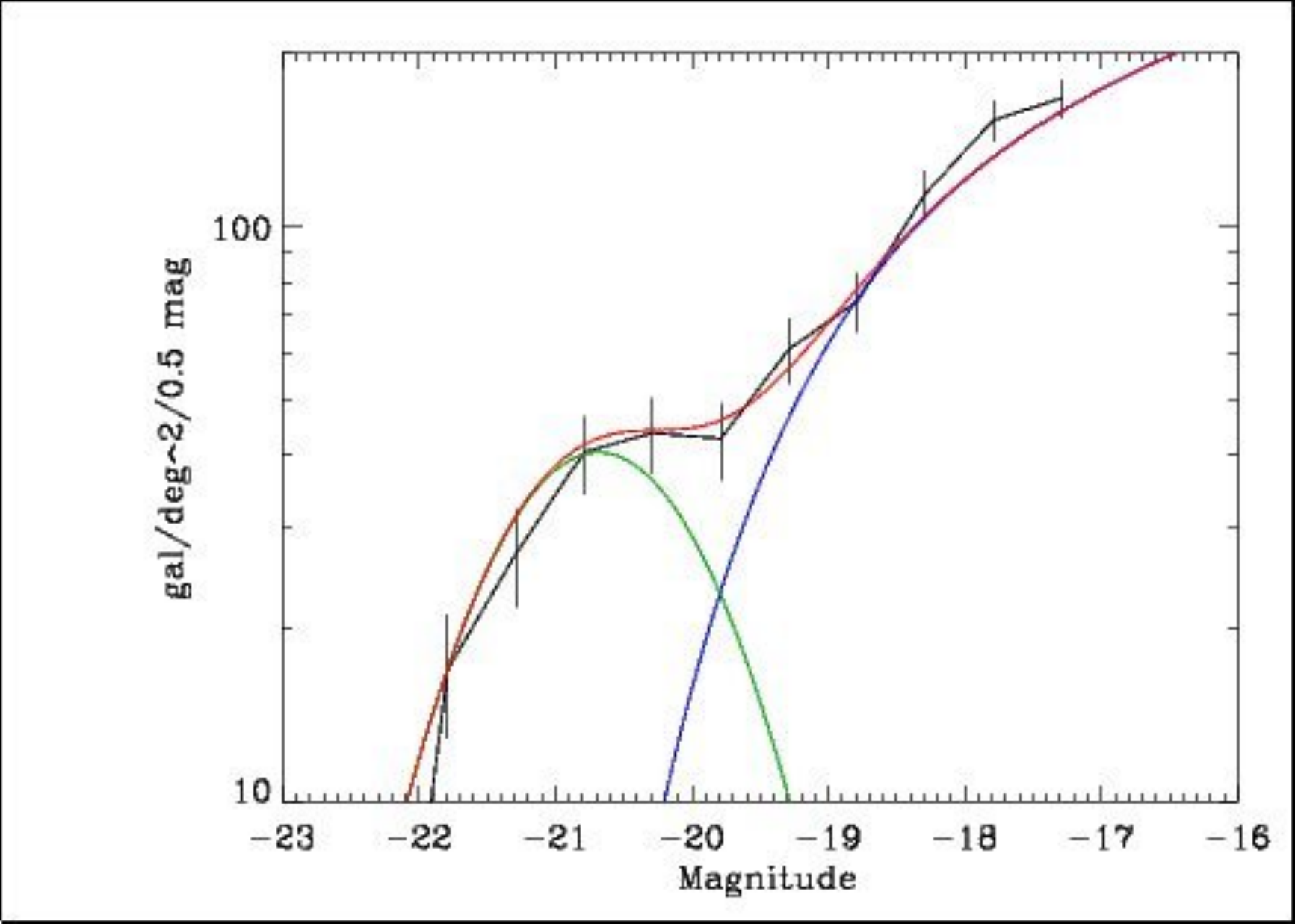}
\includegraphics[width=4cm, angle=0]{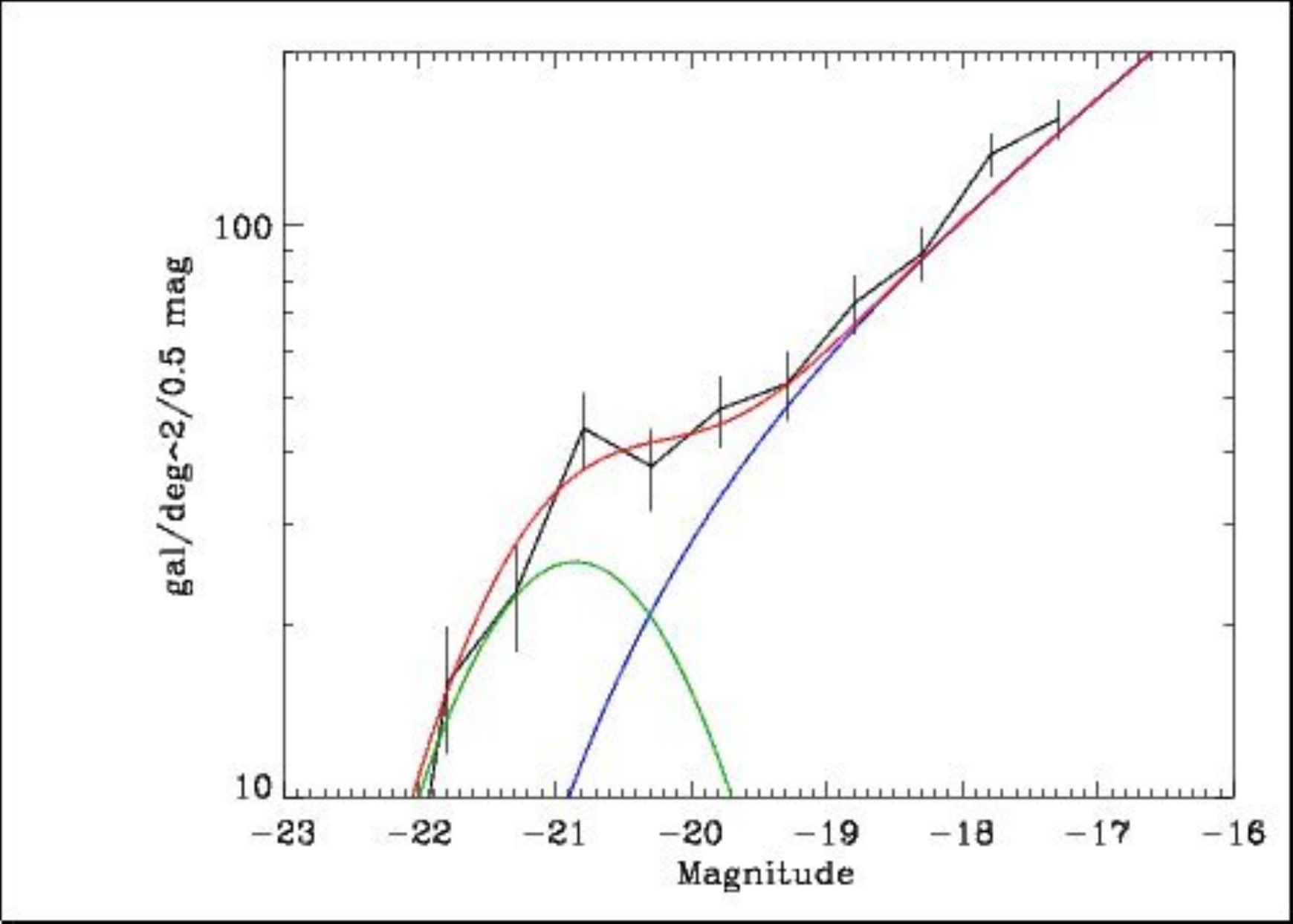}\kern0.2cm%
\includegraphics[width=4cm, angle=0]{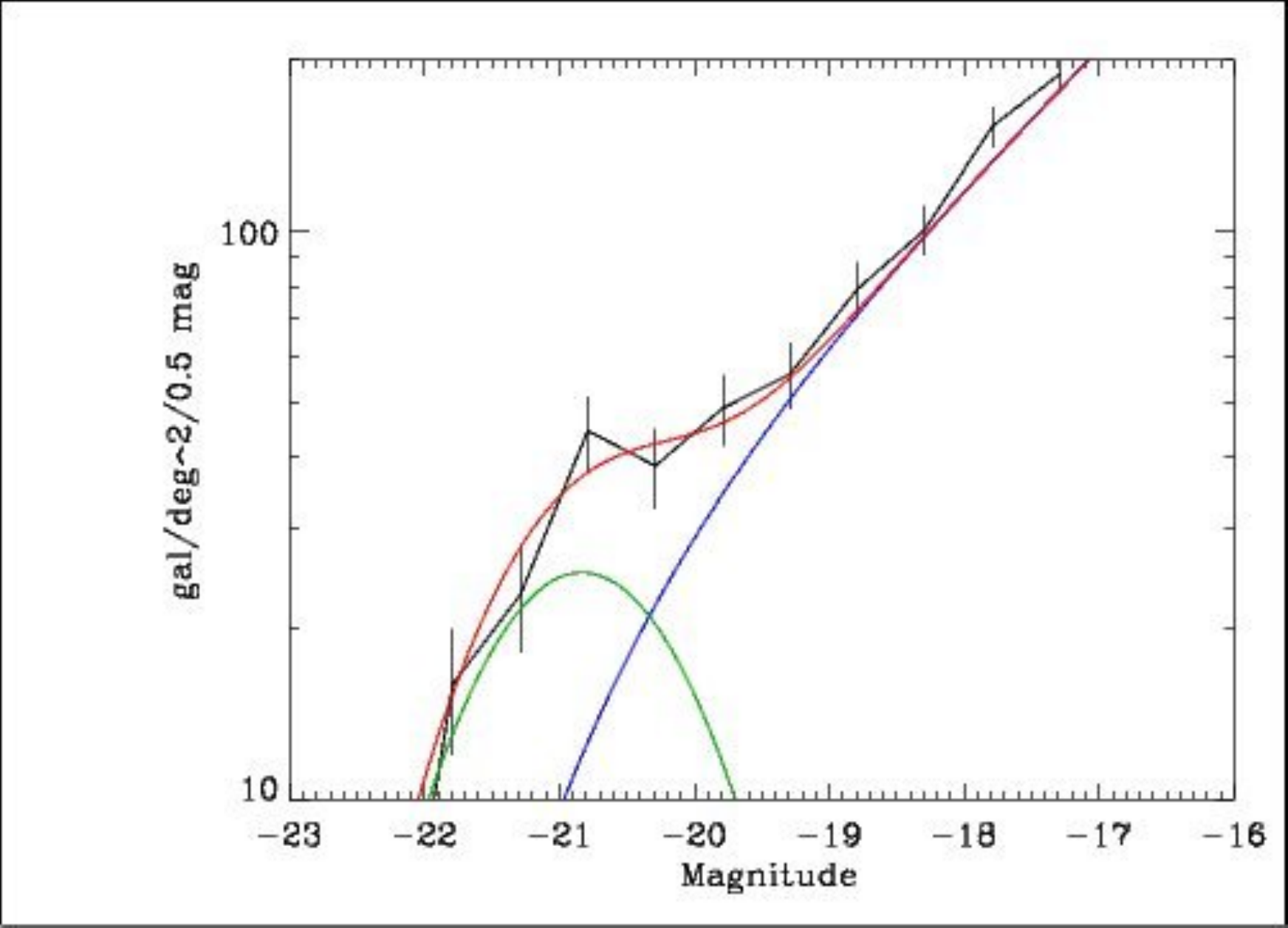}
\caption{Galaxy luminosity function of the Ophiuchus cluster in the
  entire MegaCam field.  The black line and points are the data, the
  green curve is the Gaussian component, the blue curve is the
  Schechter function, and the red curve is the total of the two
  components. Top: constant galactic extinction correction of 1.357,
  bottom: individual galactic extinction correction. Left:
  subtraction of Yasuda background counts, and right: subtraction of
  Yasuda background counts multiplied by $f=0.7$ (see Section~3.1).}
\label{fig:GLF}
\end{center}
\end{figure}

To compute the GLF of the Ophiuchus cluster, it is necessary to
subtract statistically the background contribution. Our original
galaxy counts in the CFHT/MegaCam field are shown in
Fig.~\ref{fig:histomag}. We first corrected them for a constant
  extinction of 1.357.  Background counts were taken from Yasuda et
al. (2001), who estimated field galaxy counts from the SDSS at
relatively bright magnitudes.  They show galaxy counts in the $r'$
magnitude in their figure~4. Since these counts are taken in the same
filter as our data, in bins of 0.5~mag, and normalized to 1~deg$^2$,
we can subtract them from our counts in a straightforward way to
obtain the GLF (after normalizing our data to the area covered by our
image: 0.9882~deg$^2$).  As a check to the Yasuda et al. counts, we
also plotted on the same figure the VVDS counts by McCracken et
al. (2003), that we cannot use here since they start at a magnitude of
18. We can see in Fig.~\ref{fig:counts} that the two sets of
background counts match well. We converted apparent to absolute
magnitudes by applying the distance modulus of 35.54, and thus
obtained Fig.~\ref{fig:counts}. In view of the small redshift of the
cluster, we applied no k-correction.

However, we can see in Fig.~\ref{fig:counts} that the Yasuda
background counts become larger than our galaxy counts for
$M_{r'}\sim -16.3$, corresponding to $r'\sim 19.2$, a value which is
more than one magnitude brighter than our completeness limit
($r'=20.25$).  This leads us to think that due to cosmic variance the
Yasuda counts are an overestimate of the background counts in the
direction of Ophiuchus. Our extensive redshift catalogue in this area
(Wakamatsu et al. in preparation, see also Wakamatsu et al., 2005)
indeed shows that besides the foreground (cz$< 6000$~km~s$^{-1}$)
local void, there is also a large background void behind the Ophiuchus
cluster, up to cz$\sim$16000~km~s$^{-1}$.  Therefore the background
contamination for the cluster should mostly come from the background
galaxies having velocities cz$>$16000~km~s$^{-1}$.

To examine how Yasuda's background data are modified by this cosmic
variance, we carried out numerical simulations by changing the values
of various parameters of the Schechter function, such as $M^*$,
$\alpha$, and the faintest magnitude cutoff, as well as the
integration limit of the depth (distance modulus $(m-M) <40.5$).  The
results show that these voids affect effectively the GLF parameters
for $r'>17$, or $M_{r'}>-17.5$ in absolute magnitude.  However, we
cannot obtain a unique solution, and all we can say is that the Yasuda
background counts should be multiplied by a factor $f= 0.85\pm
0.15$.
We will therefore fit the various GLFs with the values of $f$ that
bracket this interval: $f=1.0$ and $f=0.70$, noted respectively $f$ and 
$f'$ hereafter.

Since our galaxy counts show an excess of bright galaxies, we
  accounted this excess with a Gaussian function $G(M)$ (the values of
  the reduced $\chi ^2$ are indeed much lower when a Gaussian is
  included), and we fit the GLFs with a global function:
$$ GLF(M)= S(M)+G(M) $$
Here $S(M)$ is the Schechter function defined in terms of the
absolute magnitude $M$ as:
$$ S(M) = 0.4 \, \ln 10 \, \Phi^{\ast} \, y^{\alpha+1} \, e^{-y} $$
with $y=10^{0.4 \, (M^{\ast}-M)} $, where $\Phi^{\ast}$ is a
normalisation factor, $M^{\ast}$ is the typical magnitude separating
the bright and faint regimes of the GLF, and $\alpha$ is the faint end
slope.  
$G(M)$ is a Gaussian function to account for the bright part:
$$ G(M) = A\ exp[ (-4 * ln(2) * (M-M_c)^2 ) / (fwhm^2) ] $$
where $M_c$ is the central magnitude, $fwhm$ is the full-width-at-half
maximum and $A$ is the amplitude for $M=M_c$. 

We also tried to fit the GLFs with a global function including the factor
$f\leq 1$ by which the Yasuda background counts should be multiplied:
$$ GLF(M)= S(M)+G(M)+f*B(M). $$
However, the addition of a seventh free parameter made the solution
quite uncertain: the reduced $\chi^2$ remained almost the same for all
the values of $f$, and the $f$ parameter space did not seem to be
explored properly, so it was not possible to estimate $f$ with this
method. We therefore decided to fit all the GLFs with the sum of a
Gaussian and a Schechter functions, multiplying the Yasuda counts by
$f_1$ and $f'$.

It should be noted that the large excess of bright galaxies in the GLFs
is not affected by the ambiguity of the field galaxy correction,
which is very small for bright galaxies.

\begin{table*}
\begin{center}
  \caption{Parameters of the GLF for the overall image, with
    magnitudes corrected for extinction with a constant value (ct.),
    or with their individual values (var.), as well as for the
    rectangular region having the same physical size as the GLF
    computed for the Coma cluster, and for the three circular regions
    C, NW and SE (see text). $f$ is the factor by which the Yasuda
    background counts were multiplied to account for cosmic variance.}
\begin{tabular}{l|c|cccccc}
\hline
\hline
              & $f$ & $\Phi^*$  & M$^*$          & $\alpha$        & $A$        & M$_c$          & $fwhm$   \\
\hline
Overall (ct.) & 1.0 & $179\pm 130$ & $-18.9\pm 1.0$ & $-0.97\pm 0.40$ & $42\pm 7$ & $-20.6\pm 0.3$ & $0.7\pm 0.1$ \\
Overall (var.)& 1.0 & $145\pm 137$ & $-19.2\pm 1.1$ & $-1.19\pm 0.36$ & $40\pm 9$ &$-20.7\pm 0.3$  & $0.7\pm 0.1$ \\
Overall (ct.) & 0.7 & $34\pm 115$  & $-20.8\pm 3.8$ & $-1.48\pm 0.37$ & $26\pm 34$ & $-20.9\pm 0.3$ & $0.7\pm 0.2$ \\
Overall (var.)& 0.7 & $25\pm 52$   & $-21.1\pm 2.6$ & $-1.59\pm 0.19$ & $25\pm 19$ &$-20.8\pm 0.2$  & $0.7\pm 0.2$ \\
\hline
Rectangle (ct.) & 1.0 & $317\pm 108$& $-18.7\pm 0.8$ & $-0.76\pm 0.28$ & $65\pm 10$ & $-20.5\pm 2.2$ & $1.1\pm 0.5$ \\
Rectangle (var.)& 1.0 & $177\pm 120$& $-19.7\pm 1.0$ & $-1.08\pm 0.25$ & $56\pm 18$ & $-20.8\pm 0.3$ & $0.7\pm 0.1$ \\
Rectangle (ct.) & 0.7 & $217\pm 268$& $-19.4\pm 2.0$ & $-1.07\pm 0.51$ & $58\pm 25$ & $-20.9\pm 0.8$ & $0.9\pm 0.5$ \\
Rectangle (var.)& 0.7 & $ 90\pm  78$& $-20.6\pm 1.3$ & $-1.32\pm 0.16$ & $41\pm 25$ & $-20.9\pm 0.2$ & $0.6\pm 0.1$ \\
\hline
C (ct.)       &1.0 & $428\pm 127$&$-19.4\pm 0.4$  &$-0.81\pm 0.22$  &$80\pm 4$   &$-21.3\pm 0.1$  &$0.7\pm 0.2$   \\
C (var.)      &1.0 & $551\pm 112$&$-18.8\pm 0.4$  &$-0.58\pm 0.23$  &$76\pm 6$   &$-21.0\pm 0.4$  &$1.4\pm 1.0$   \\
C (ct.)       &0.7 & $435\pm  41$&$-19.4\pm 0.2$  &$-0.83\pm 0.04$  &$86\pm 8$   &$-21.3\pm 0.1$  &$0.6\pm 0.1$   \\
C (var.)      &0.7 & $437\pm  49$&$-19.4\pm 0.2$  &$-0.83\pm 0.05$  &$76\pm 7$   &$-21.3\pm 0.2$  &$0.8\pm 0.2$   \\
\hline
NW (ct.)      & 1.0 & $2\pm 4$ & $-34.5\pm 7.9$ & $-1.28\pm 0.04$ & & & \\
NW (var.)     & 1.0 & $4\pm 6$ & $-31.4\pm 5.3$ & $-1.28\pm 0.02$ & & & \\
NW (ct.)      & 0.7 & $82\pm 54$ & $-19.8\pm 0.8$ & $-1.35\pm 0.15$ & $81\pm 10$ & $-20.8\pm 0.1$ & $0.6\pm 0.1$ \\
NW (var.)     & 0.7 & $379\pm 62$ & $-17.5\pm 0.3$ & $-0.47\pm 0.24$ & $81\pm 6$ & $-20.2\pm 0.2$ & $1.5\pm 0.3$ \\
\hline
SE (ct.)      & 1.0 & $16\pm 14$ & $-22.9\pm 1.3$ & $-1.51\pm 0.07$ & & & \\
SE (var.)     & 1.0 & $0.6\pm 1.0$ & $-31.9\pm 4.4$ & $-1.43\pm 0.02$ & & & \\
SE (ct.)      & 0.7 & $537\pm 57$ & $-17.4\pm 0.2$ & $0.42\pm 0.34$ & $74\pm 6$ & $-20.3\pm 0.1$ & $0.8\pm 0.1$ \\
SE (var.)     & 0.7 & $573\pm 63$ & $-17.8\pm 0.3$ & $-0.31\pm 0.31$ & $91\pm 7$ & $-20.3\pm 0.1$ & $0.6\pm 0.1$ \\
\hline                                       
\end{tabular}
\label{tab:GLFparam}
\end{center}
\end{table*}

We computed the GLF both for a constant (noted ``ct.'' in
Table~\ref{tab:GLFparam}) Galactic extinction of 1.357 and for a
Galactic extinction that is different for each galaxy (noted ``var.''
in Table~\ref{tab:GLFparam}), as explained above. The results of the
Gauss+Schechter fits are shown in Fig.~\ref{fig:GLF} and the
corresponding parameters are given in Table~\ref{tab:GLFparam}. The
reduced $\chi ^2$ values of the fits are 0.81 and 0.72 for a
constant and a variable extinction correction respectively.  The
reduced $\chi ^2$ values are larger than 2.0 if no Gaussian component
is included, therefore justifying our choice of a Gauss+Schechter fit
to the GLFs.

We can see that the galaxy counts do not vary strongly with the method
used to correct for Galactic extinction (Fig.~\ref{fig:counts}), as
confirmed by the fact that the shapes and parameters of the GLF fits
are quite similar in both cases, and agree within the error bars
(Table~\ref{tab:GLFparam}). Of course if we take $f'=0.7$ the
  counts are higher, and the parameters of the fit change. 

We can note that the faintest bins in which we can compute the GLF are
at absolute magnitudes of $-17.5$ and $-16.5$, for $f=1.0$
  and $f'=0.7$ respectively, brighter or comparable to the
completeness limit of our photometric catalogue ($-16.4$) given in the
previous Section, so they can be considered as reliable.

The fits appear quite satisfactory in view of all the difficulties
overcome to obtain a photometric catalogue.  The Gaussian components
are in both cases important at bright magnitudes.  An excess at very
bright magnitudes seems to be quite a general phenomenon, as already
noted for example in Abell~223 (Durret et al. 2010), Abell~1758
(Durret et al. 2011) or Abell~3376 (Durret et al. 2013).  But all
these clusters were mergers, so we did not expect to find such a large
gaussian component in Ophiuchus, which is believed from previous
studies to be quite relaxed.  The faint end slope is moderate for
  $f=1.0$ ($\alpha=-0.97$ or $-1.19$, depending on the way the
  extinction is corrected), and steeper for $f'=0.7$ ($-1.48$ to
  $-1.59$).  

 The latter steep values could be due to a non negligible background
contamination at faint magnitudes. However we can note that comparable
values have already been observed in other clusters. For example, in
the nearby massive cluster Coma, Adami et al. (2007) found a faint end
slope even steeper than $-1.49$ in the north part of Coma, which
is a relatively quiescent region, and a flatter slope around $-1.3$ in
the south half of Coma, which is experiencing infalls.

This led us to derive the GLF in a region having the same physical
size as for Coma, to make the comparison of these two clusters more
reliable. Our results are described in the next Section.

\subsection{The galaxy luminosity function in a physical region similar to
our previous study of the Coma GLF}
\label{subsec:compComa}

Adami et al. (2007) analysed the GLF of the Coma cluster in a region
covering 40~arcmin in right ascension and 50~arcmin in declination. At
the redshift of Coma, this corresponds to a zone of
$1.1136\times 1.3920$~Mpc$^2$. Since the scale for Ophiuchus is 0.585
kpc~arcsec$^{-1}$, this size corresponds to a region of
$0.5288\times 0.661$~deg$^2$, or a surface of 0.34954~deg$^2$.  We
therefore extracted from our photometric catalogue a zone defined by
257.8511$<$RA$<$258.3799, $-23.7003<$DEC$<-23.0393$, containing  1248 galaxies.

\begin{figure}
 \begin{center}
\includegraphics[width=4cm, angle=0]{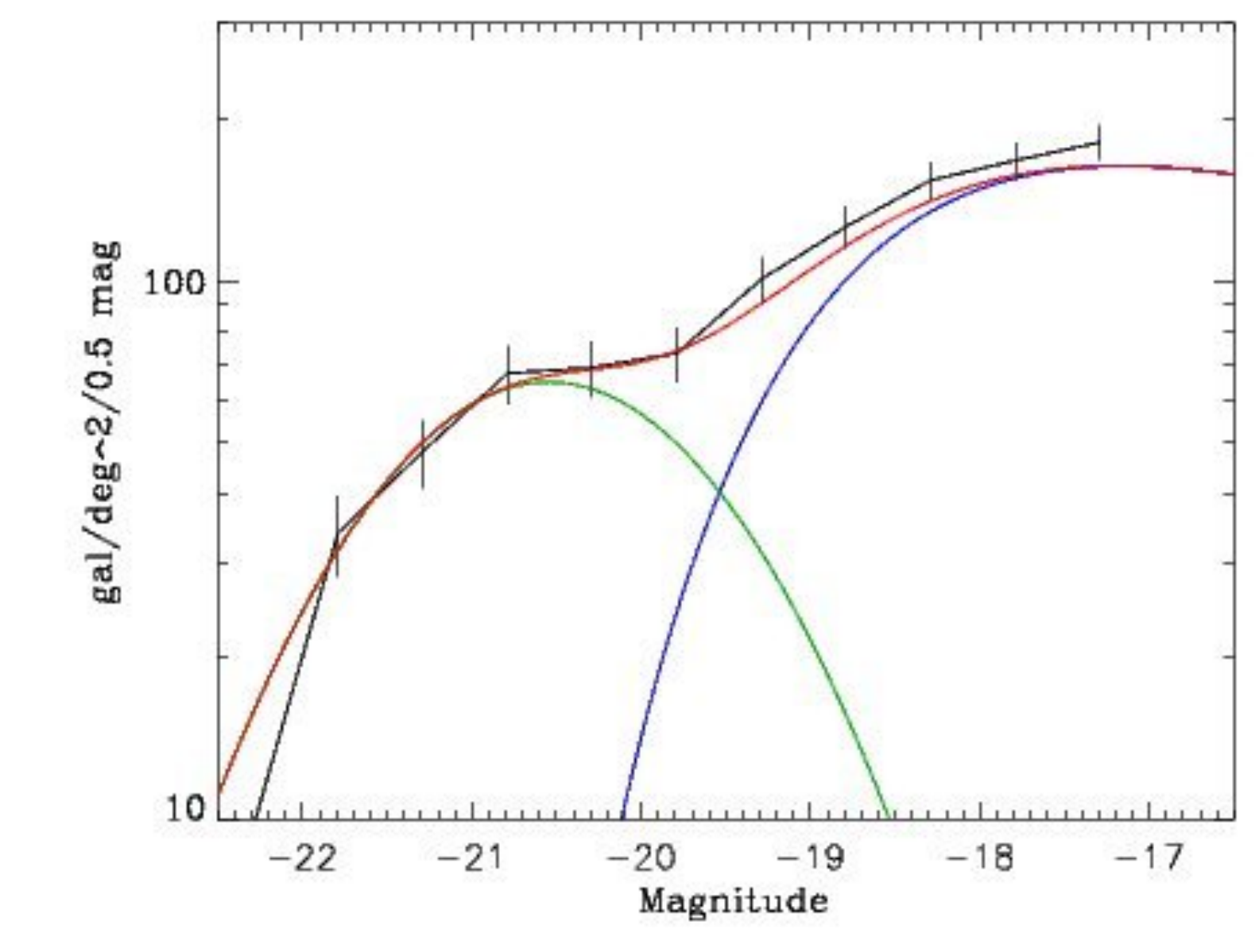}\kern0.2cm%
\includegraphics[width=4cm, angle=0]{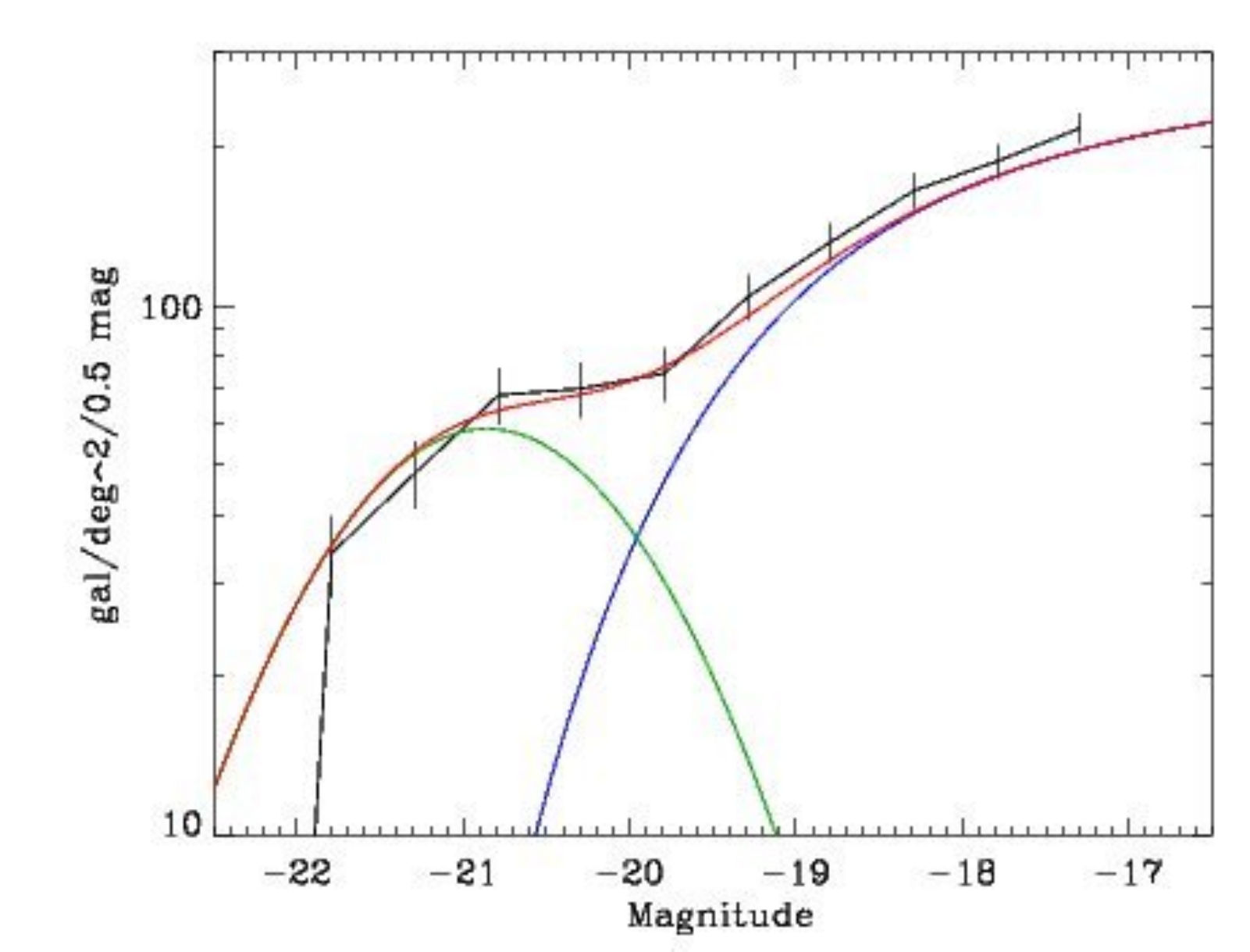}
\includegraphics[width=4cm, angle=0]{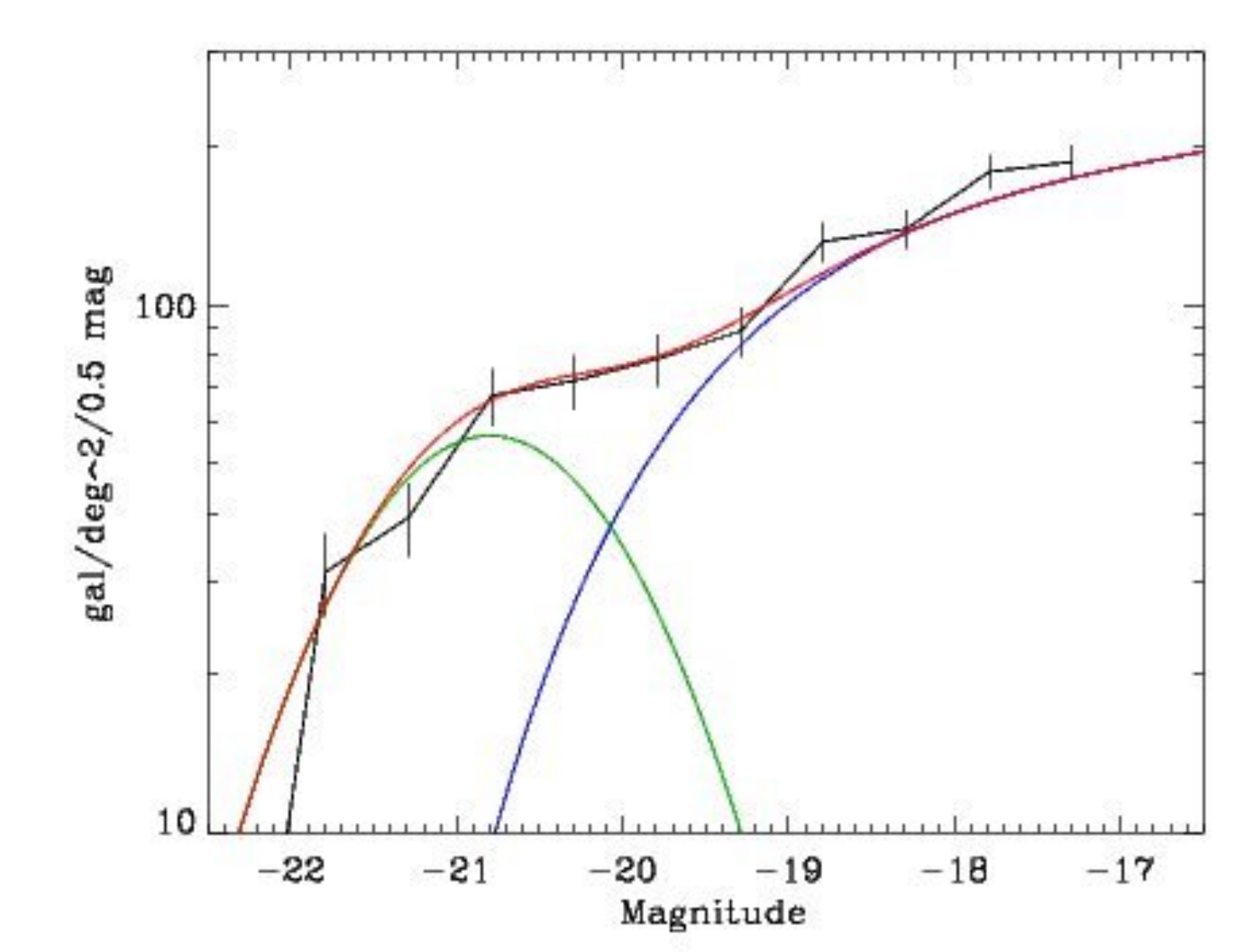}\kern0.2cm%
\includegraphics[width=4cm, angle=0]{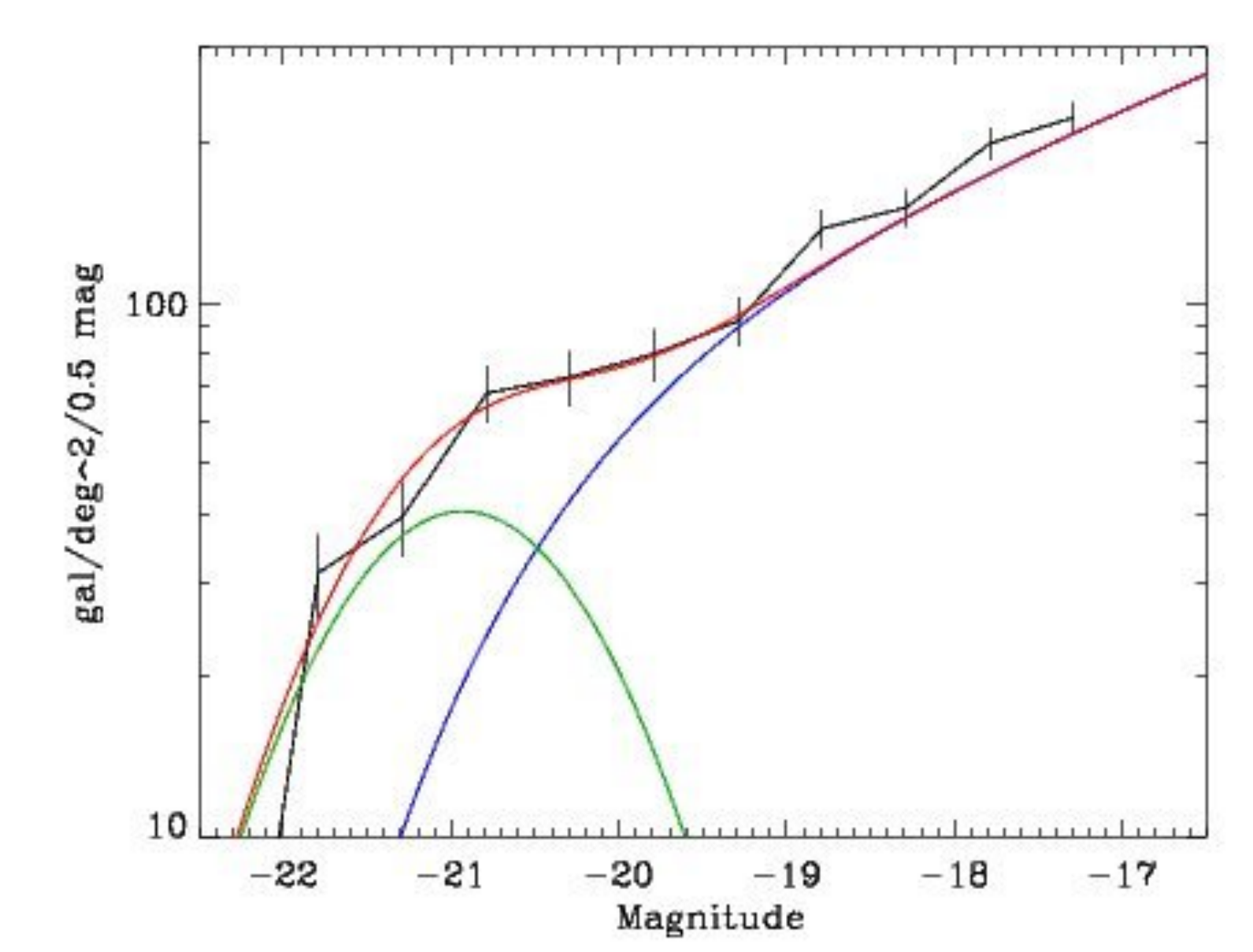}
\caption{Galaxy luminosity function of the Ophiuchus cluster in the
  ``rectangle'' region (covering the same physical size as for the
  Coma cluster, see text). Top: constant galactic extinction, bottom:
  individual galactic extinction correction. Left: subtraction
    of Yasuda background counts, and right: subtraction of Yasuda
    background counts multiplied by $f=0.7$ (see Section~3.1). See
  caption of Fig.~\ref{fig:GLF} for details.}
\label{fig:GLF_Coma}
\end{center}
\end{figure}

We computed the GLF in the same way as described above and fit a
Gaussian and a Schechter functions. The corresponding GLFs and their
fits are shown in Fig.~\ref{fig:GLF_Coma} for two cases of different
absorption corrections and the best fit parameters are given in
Table~\ref{tab:GLFparam} (region noted ``rectangle'').

We will compare these GLFs with that obtained for Coma by Adami et
al. (2007) in Section~\ref{sec:concl}.

\subsection{Defining regions in Ophiuchus through density maps}
\label{subsec:densmaps}

\begin{figure}
 \begin{center}
\includegraphics[width=5.2cm, angle=0]{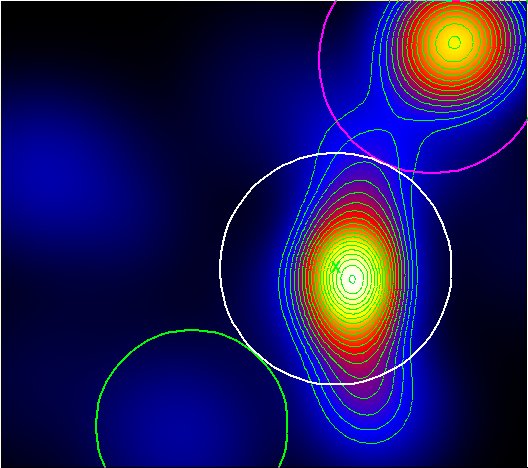}
\includegraphics[width=5.2cm, angle=0]{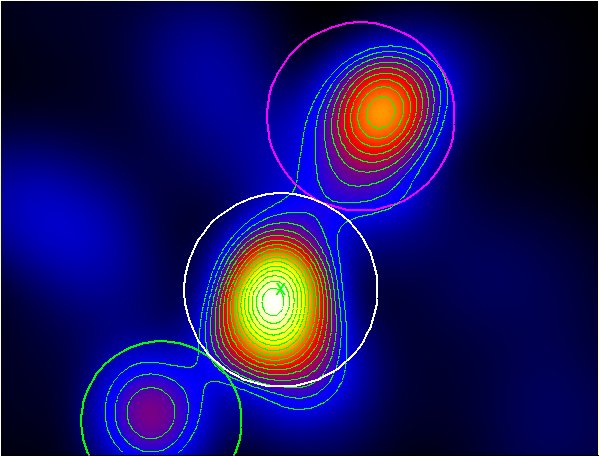}
\includegraphics[width=5.2cm, angle=0]{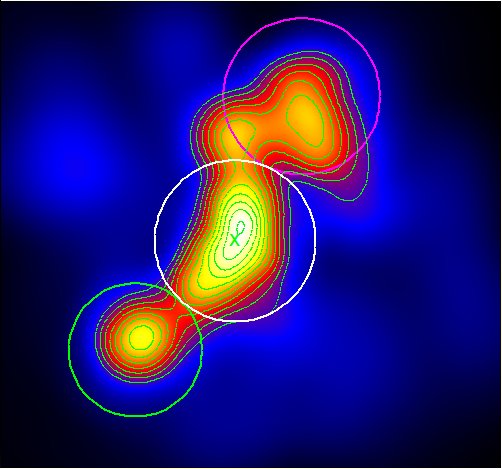}
\includegraphics[width=5.2cm, angle=0]{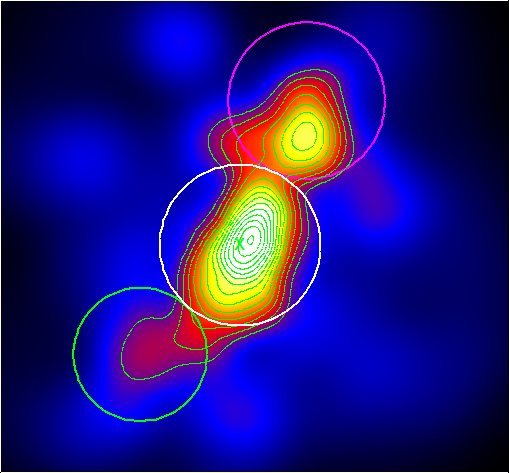}
\includegraphics[width=5.2cm, angle=0]{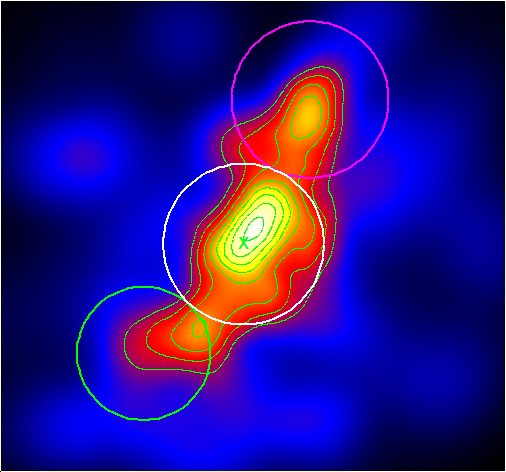}
\caption{Galaxy density maps with various magnitude limits (chosen
  arbitrarily, for magnitudes without any extinction correction). From
  top to bottom: $r<15.5$ (25 galaxies), $r<16$ (58 galaxies),
    $r<17$ (141 galaxies), $r<18$ (271 galaxies), and $r<19$ (517
    galaxies). Contour levels start at $3\sigma$ above the background
    (see text) and are spaced by $1\sigma$. The cross indicates the
    position of the cD galaxy.}
\label{fig:densmaps}
\end{center}
\end{figure}

GLFs have been observed to vary with the position within the cluster
(e.g. Adami et al. 2007 and references therein).  In order to define
the regions where it would be interesting to derive GLFs, we computed
galaxy density maps based on an adaptive kernel technique with a
  generalized Epanechnikov kernel as suggested by Silverman (1986).
  Our method is based on an earlier version developed by Timothy Beers
  (ADAPT2) and further improved by Biviano et al. (1996).  The
  statistical significance is established by bootstrap resampling of
  the data. A density map is computed for each new realisation of the
  distribution. We choose a pixel size of $0.001^\circ$ (3.6~arcsec).
  For each pixel of the map, the final value is taken as the mean over
  all realisations. A mean bootstrapped map of the distribution is
  thus obtained. The number of bootstraps used here is 100.

  The significance level of our detections was estimated from the mean
  value and dispersion of the background of each image. To estimate
  these quantities, we draw for each density map the histogram of the
  pixel intensities. We apply a 2.5$\sigma$ clipping to eliminate the
  pixels of the image that have high values and correspond to objects
  in the image. We then redraw the histogram of the pixel intensities
  after clipping and fit this distribution with a Gaussian.  The mean
  value of the Gaussian gives the mean background level, and the width
  of the Gaussian gives the dispersion, that we will call $\sigma$. We
  then compute the values of the contours corresponding to 3$\sigma$
  detections as the background plus 3$\sigma$. The contours shown in
  Fig.~\ref{fig:densmaps} start at 3$\sigma$ and increase by
  1$\sigma$.

The limitation here is that we only have photometry in one band, so we
cannot select galaxies along the red sequence (i.e. with a high
probability of belonging to the cluster). We therefore chose to
compute maps for different magnitude limits (chosen arbitrarily),
applied to the magnitudes before any extinction correction: $r'<15.5$
(25 galaxies), $r'<16$ (58 galaxies), $r'<17$ (141 galaxies), $r'<18$
(271 galaxies), and $r'<19$ (517 galaxies). Obviously, the
contamination by background galaxies becomes larger as the limiting
cut becomes fainter. 

The resulting density maps are shown in Fig.~\ref{fig:densmaps}. Since
the galaxy catalogues from which the density maps are computed have
slightly different sizes, the maps also have somewhat different
sizes. However, the radii of the three circles remain constant from
one figure to another.  

We can see that depending on the magnitude cut the aspects of the
density maps change. The cluster (white circle) is clearly the
brightest structure in all the maps. However, a structure is visible
northwest of the cluster, and a second structure appears at fainter
magnitudes southeast of the cluster.

We therefore defined three circular regions in which we derived the
GLFs: the first one (hereafter C) is centered on the cluster centre
and has a radius of 10.0~arcmin, or 351~kpc at the cluster redshift
(white circle), the northwest circle (hereafter NW) is centered on
RA=$257.9774^\circ$, DEC=$-23.0701^\circ$ and has a radius of
9.7~arcmin, or 340~kpc (magenta circle), and the southeast circle
(hereafter SE) is centered on RA=$258.3224^\circ$,
DEC=$-23.5961^\circ$ and has a radius of 8.3~arcmin, or 291~kpc (green
circle).

We note that the NW region more or less coincides with the zone
  where Watanabe et al. (2001) found an excess of X-ray emission in
  their ASCA image (see their Fig.~5), so it is probably a substructure
  in the cluster.

\subsection{The galaxy luminosity function in three regions}

\begin{figure}
 \begin{center}
\includegraphics[width=4cm,angle=0]{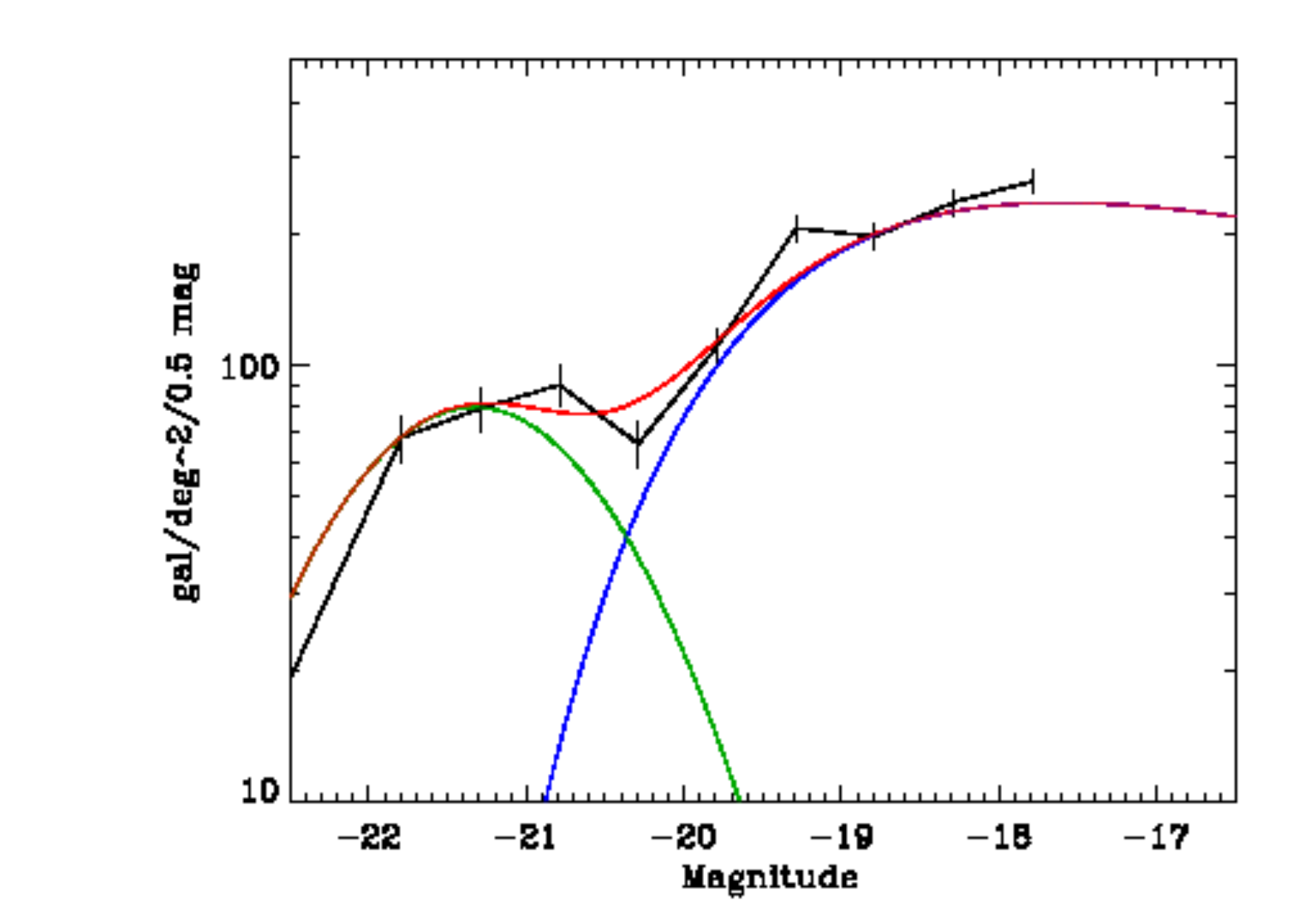}\kern0.2cm%
\includegraphics[width=4cm,angle=0]{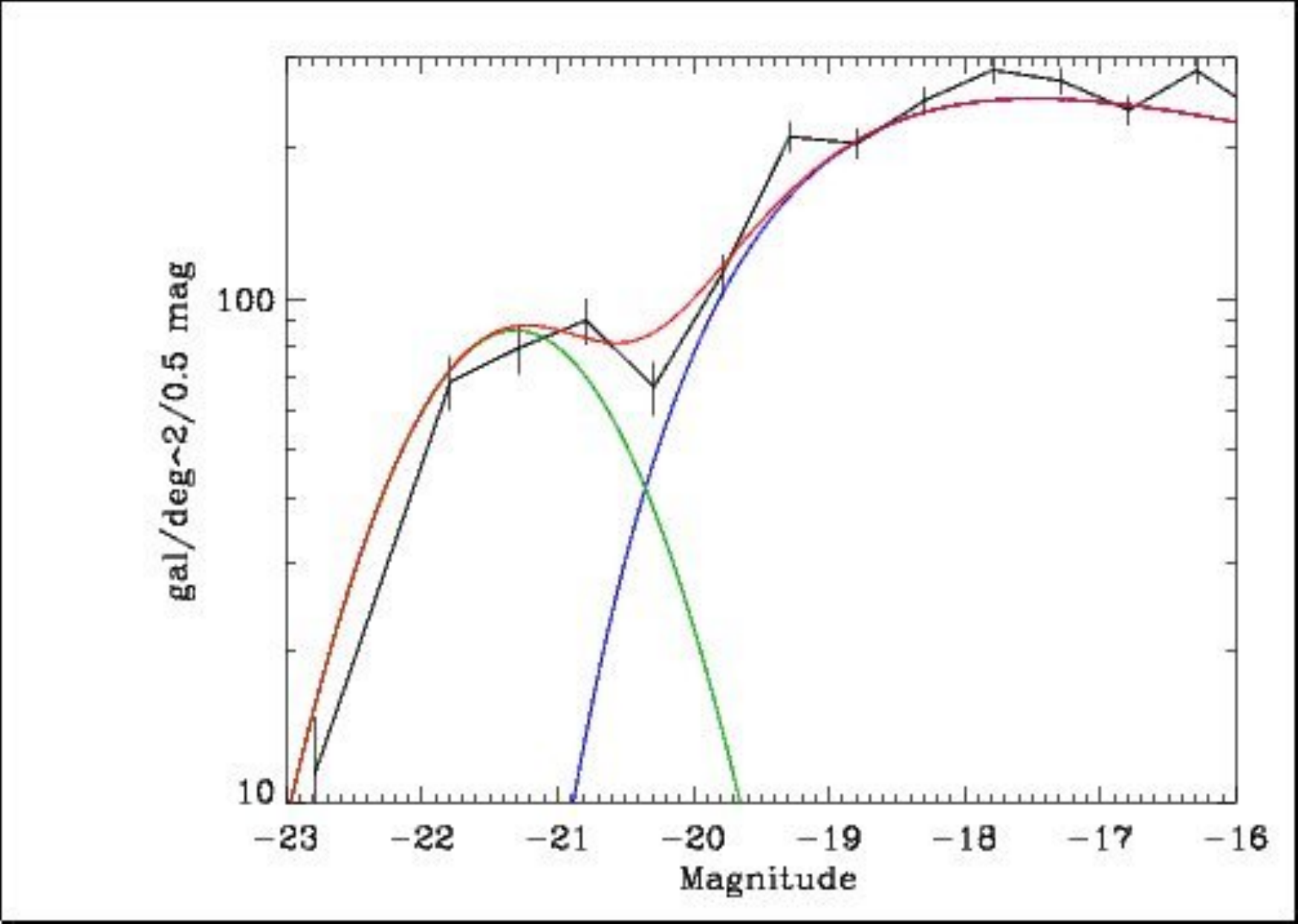}     
\includegraphics[width=4cm,angle=0]{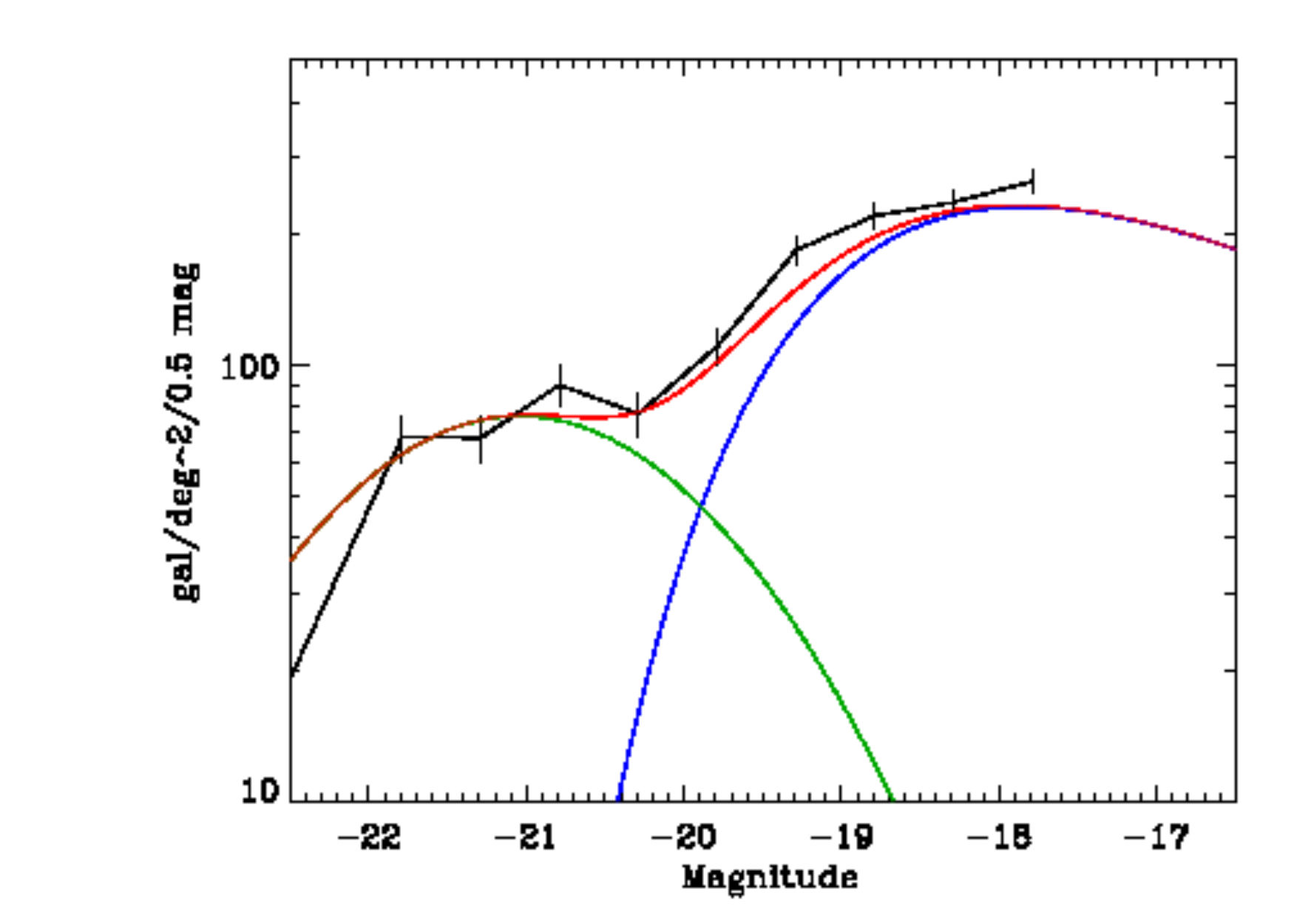}\kern0.2cm%
\includegraphics[width=4cm,angle=0]{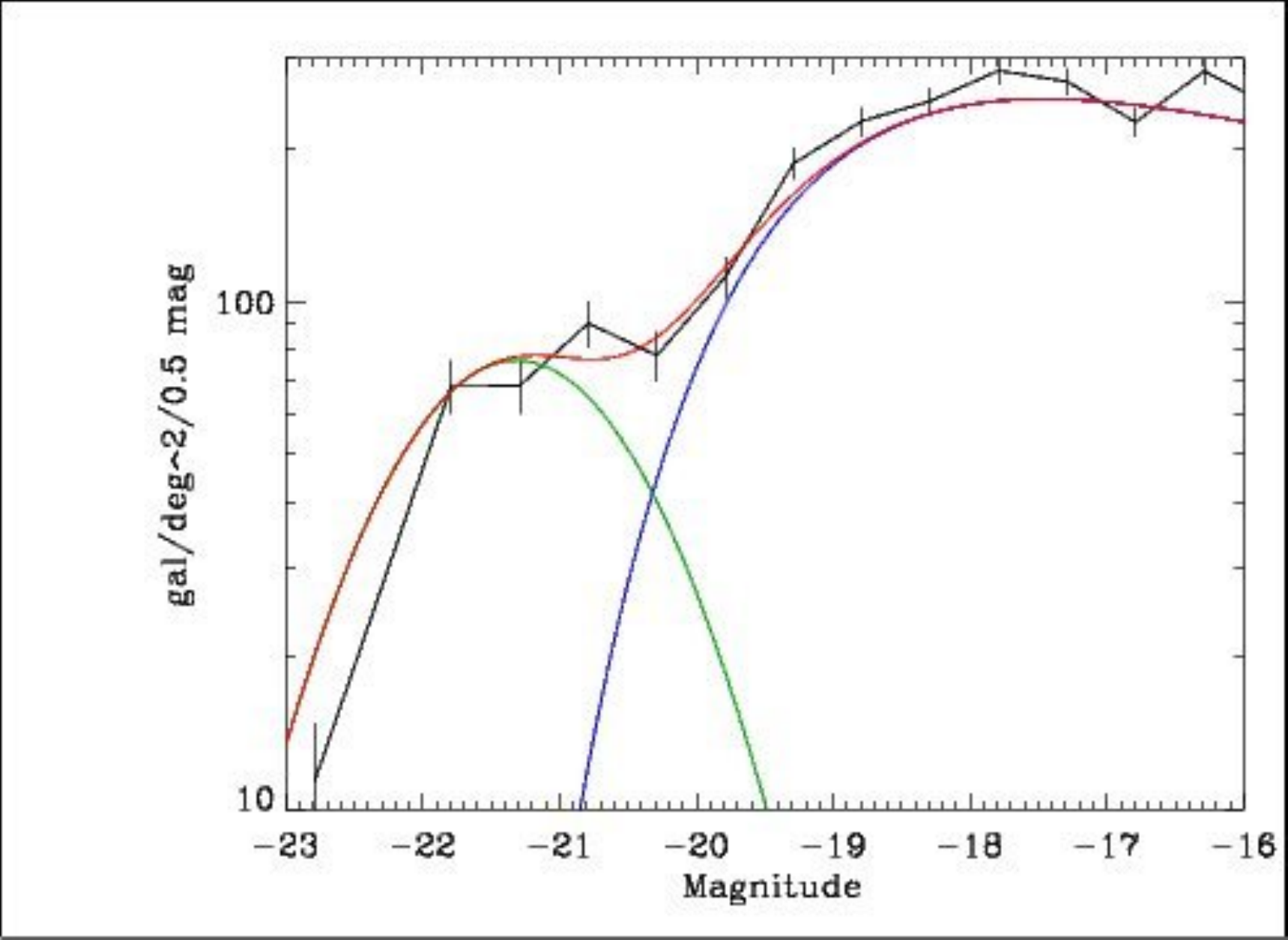}
\includegraphics[width=4cm,angle=0]{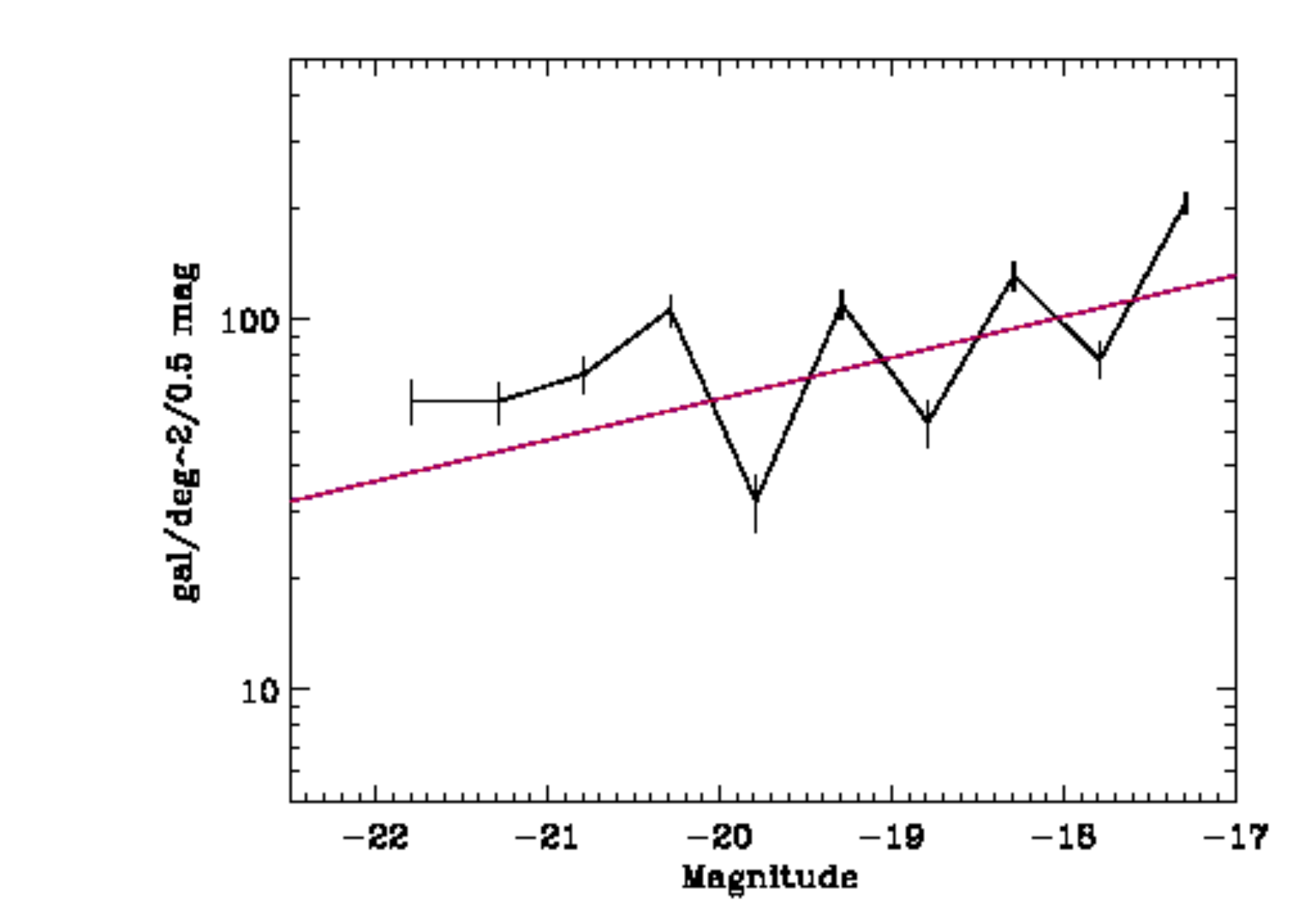}\kern0.2cm%
\includegraphics[width=4cm,angle=0]{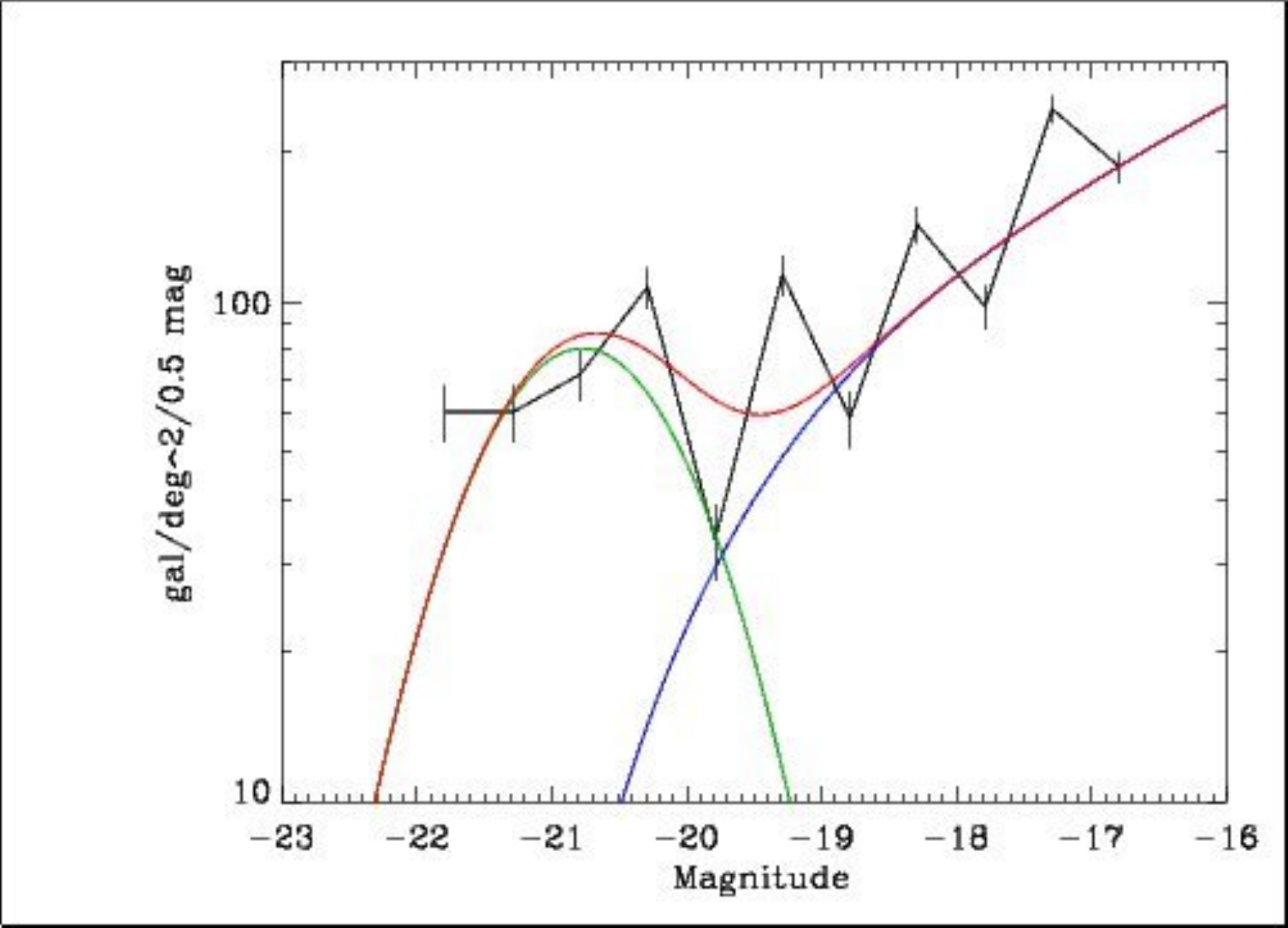}
\includegraphics[width=4cm,angle=0]{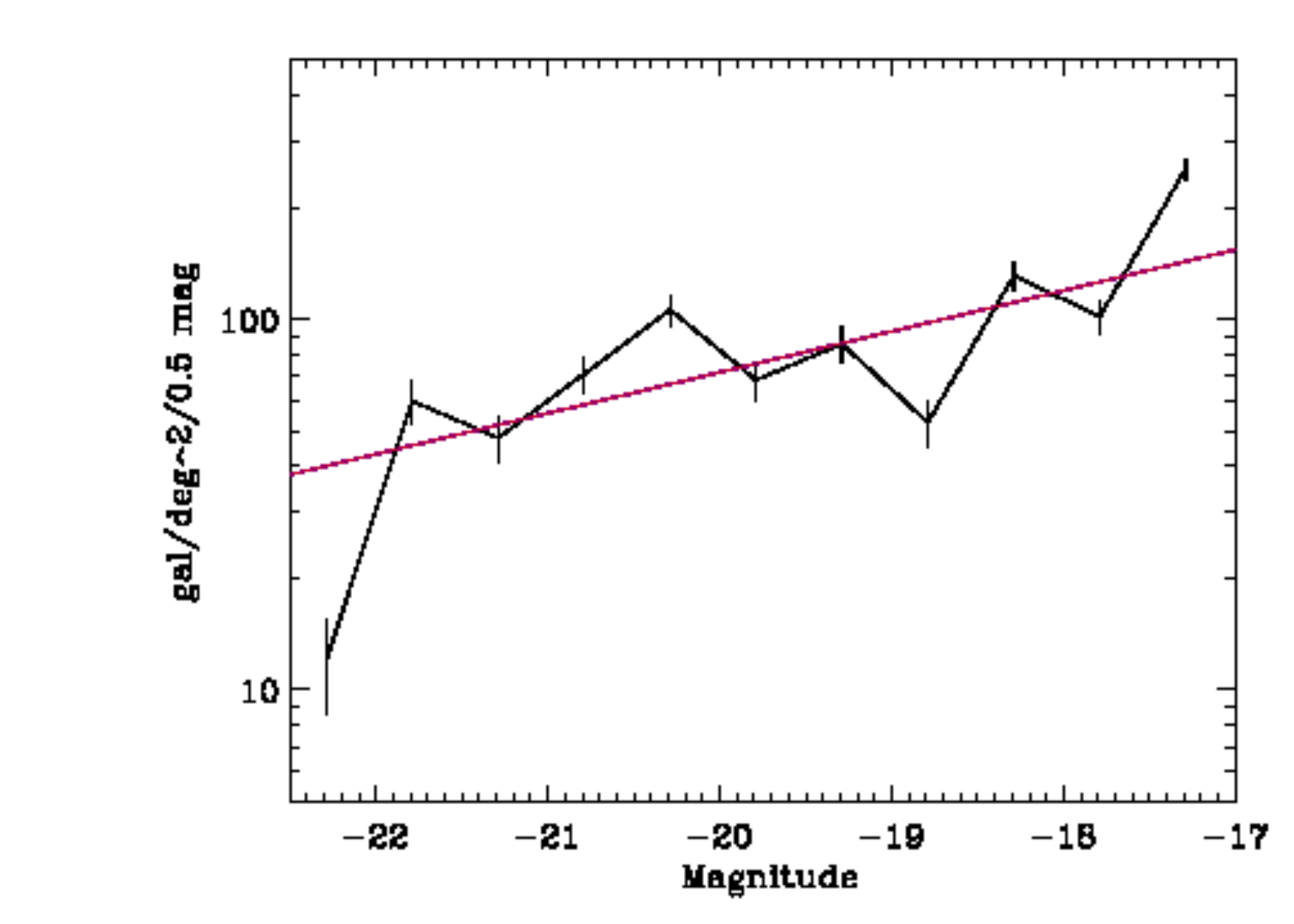}\kern0.2cm%
\includegraphics[width=4cm,angle=0]{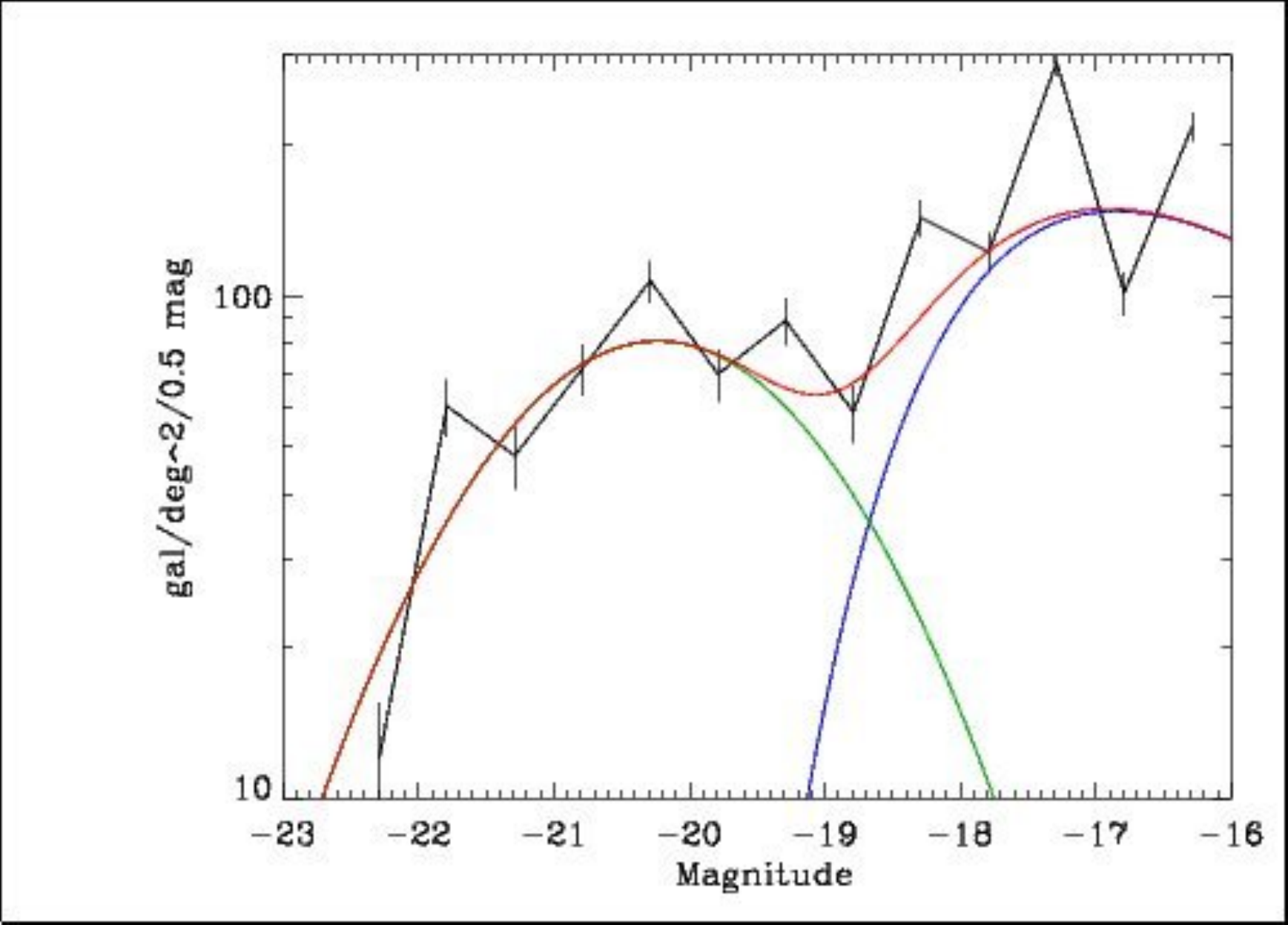}
\includegraphics[width=4cm,angle=0]{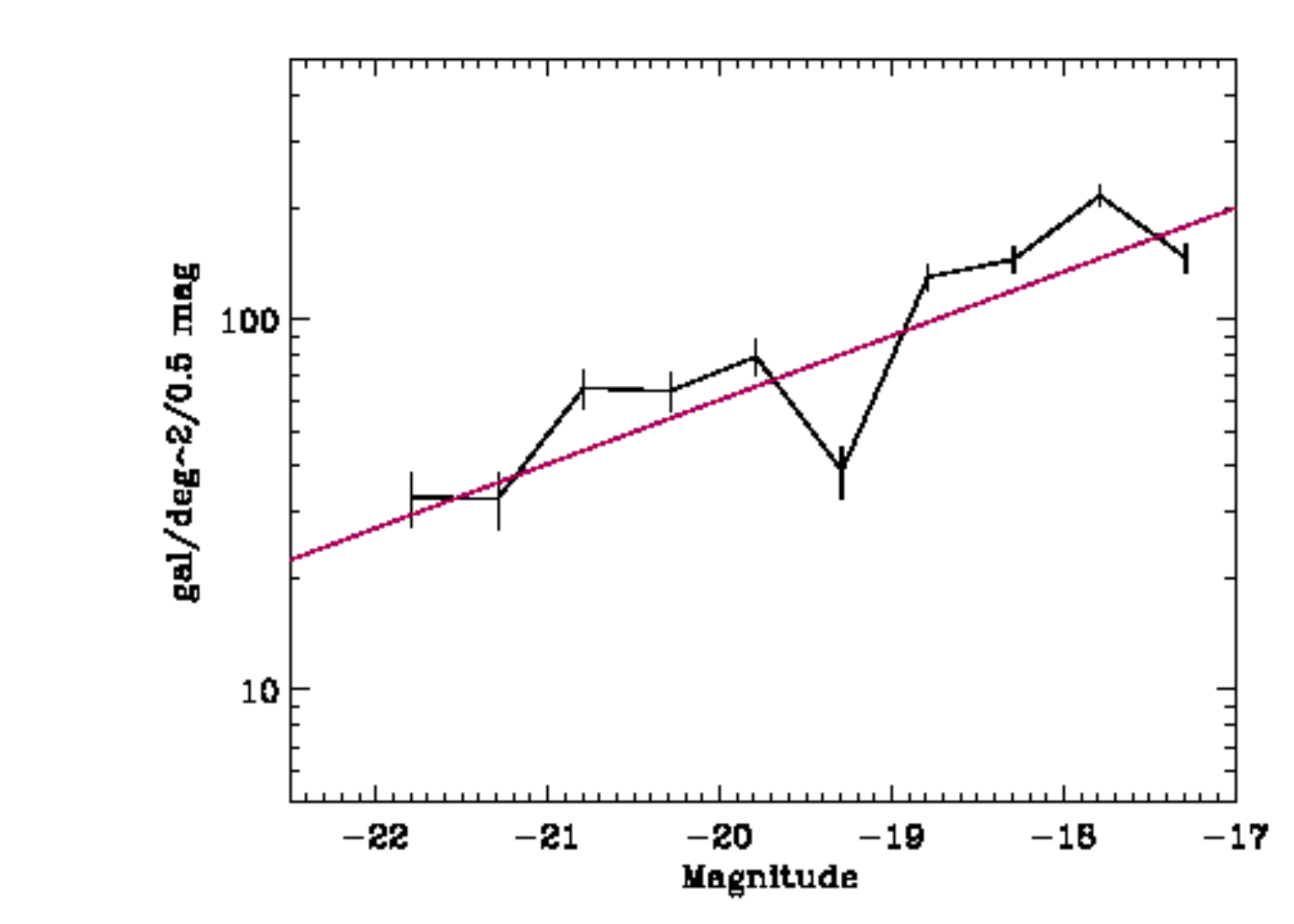}\kern0.2cm%
\includegraphics[width=4cm,angle=0]{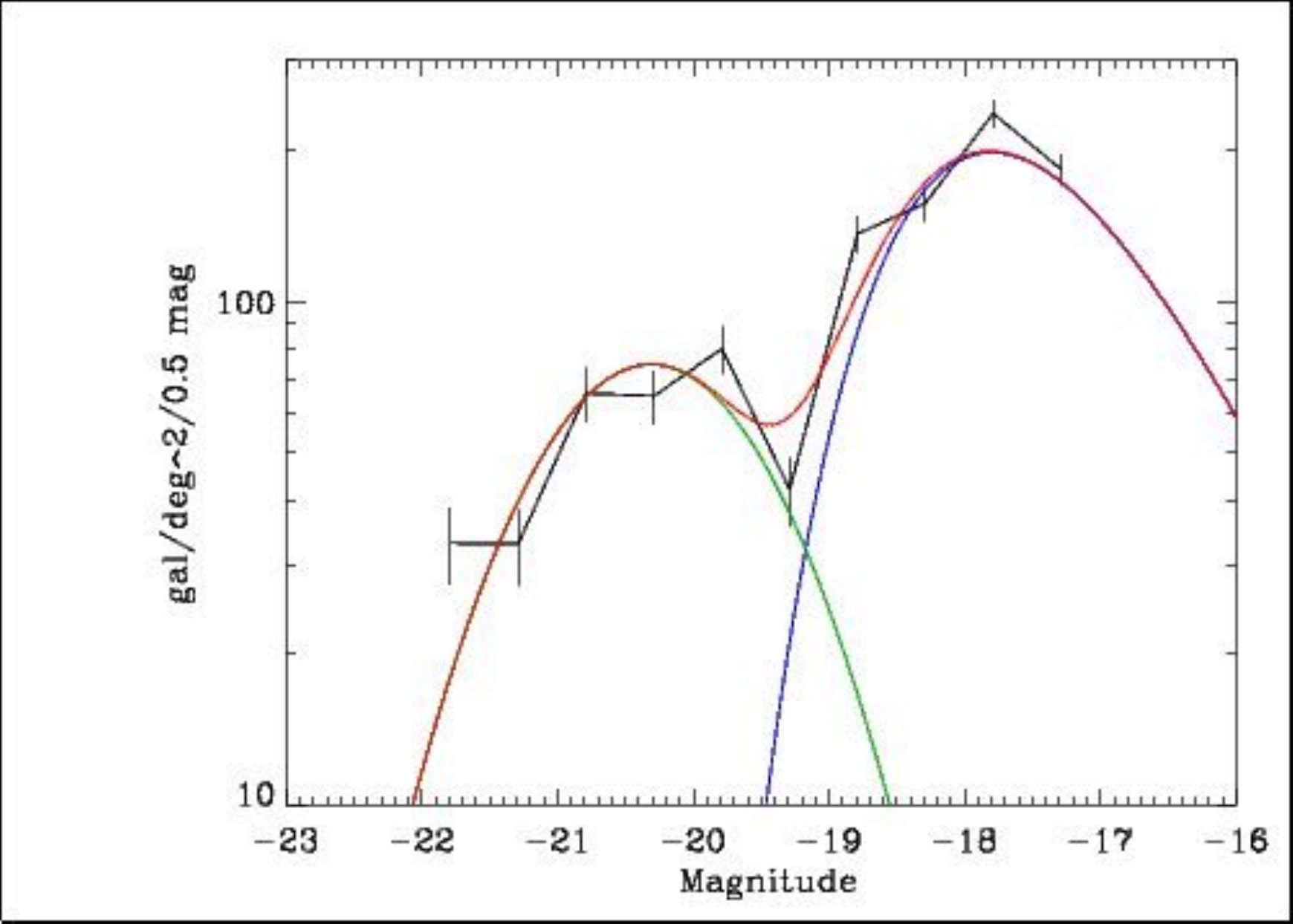}
\includegraphics[width=4cm,angle=0]{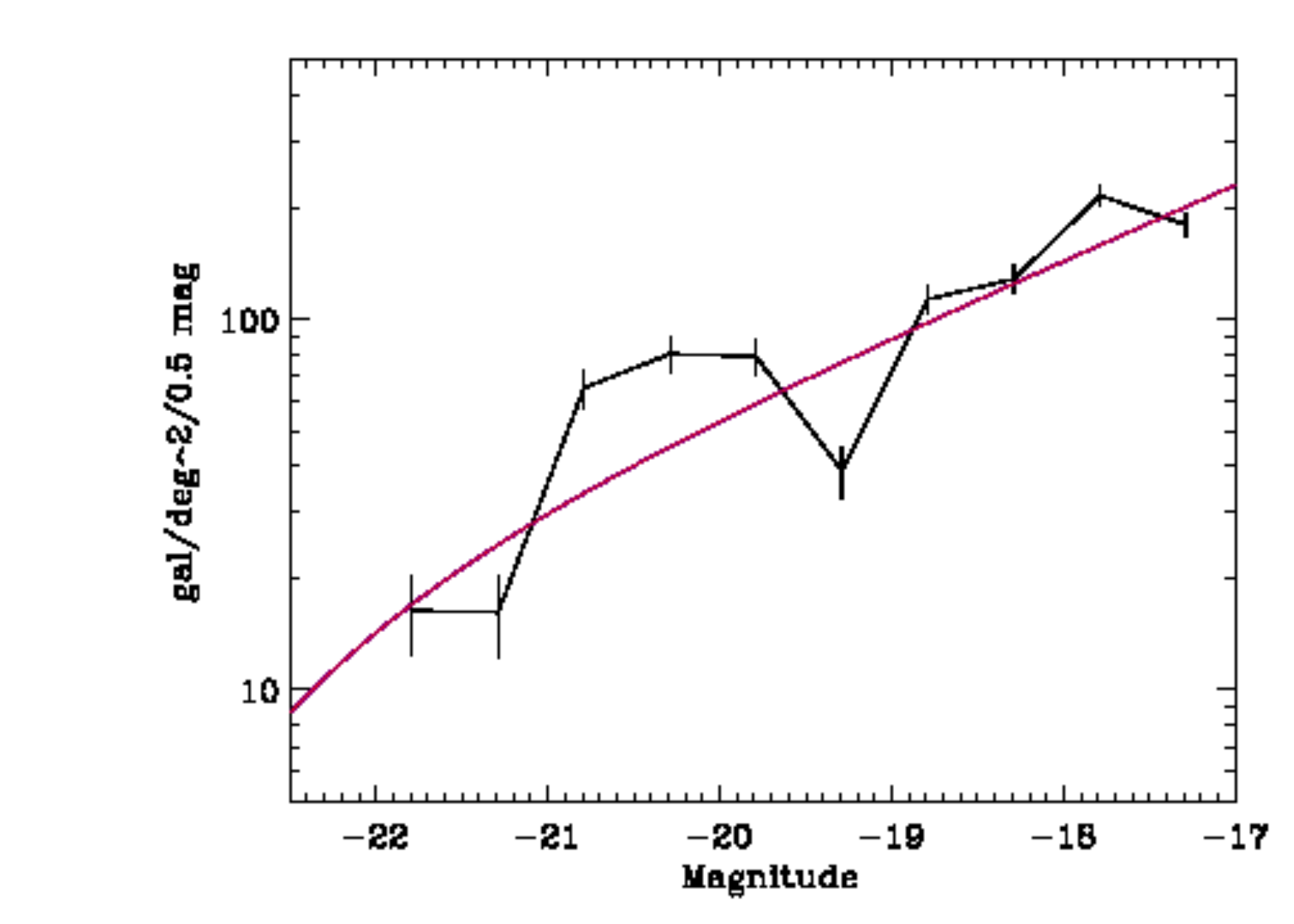}\kern0.2cm%
\includegraphics[width=4cm,angle=0]{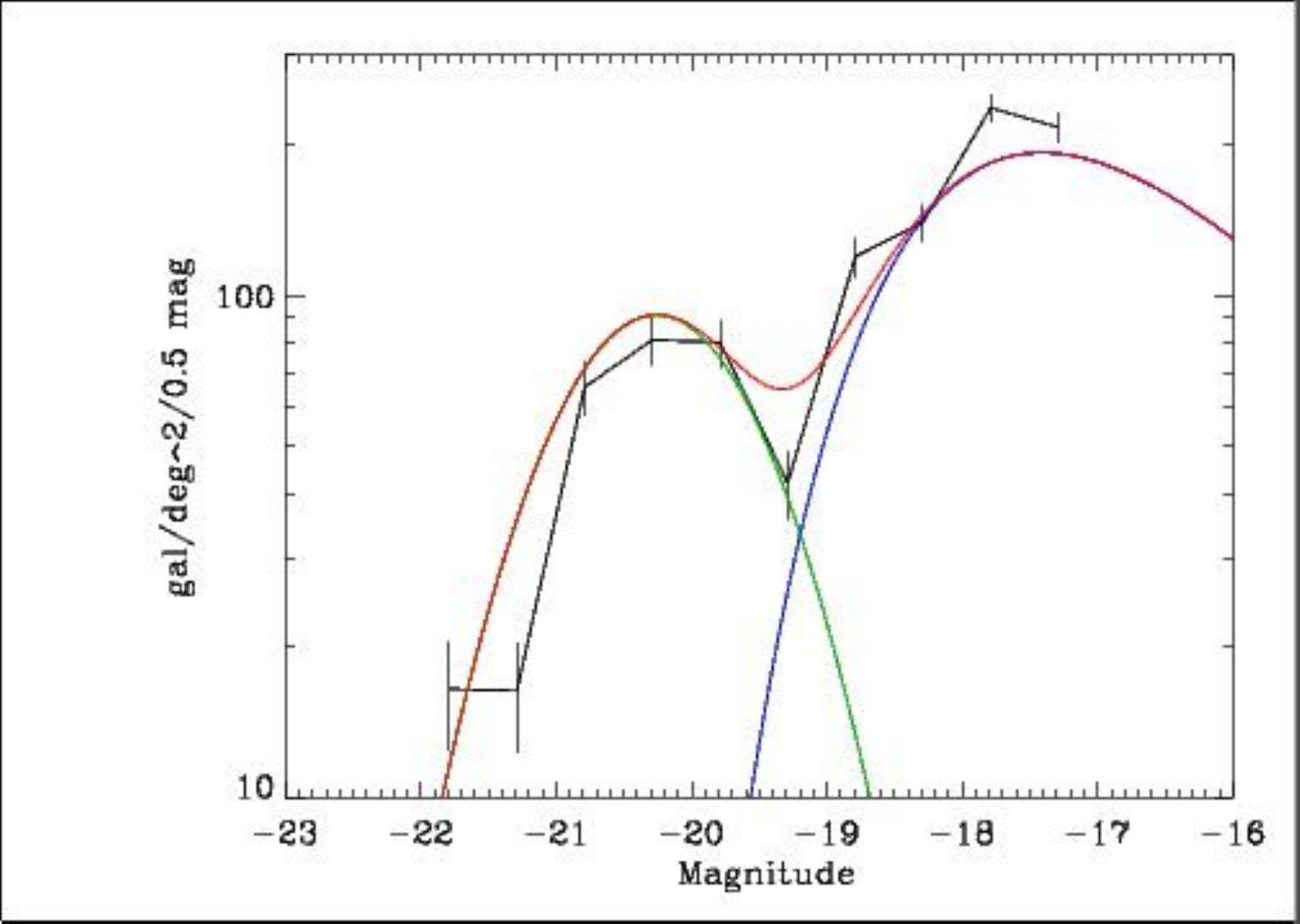}
\caption{Galaxy luminosity functions in the Centre C (two top
  figures), and in regions NW (two middle figures) and SE (two bottom
  figures) (see text). For each region, the top figure was obtained
  with a constant Galactic extinction and the bottom one with a
  different extinction correction for each galaxy. Left:
    subtraction of Yasuda background counts, and right: subtraction of
    Yasuda background counts multiplied by $f=0.7$ (see Section~3.1).
  The symbols are as in Fig.~\ref{fig:GLF}.}
\label{fig:GLFregions}
\end{center}
\end{figure}

We extracted galaxy counts in the three circular regions C, NW and SE,
and applied to each the two extinction corrections described above,
and the two possible values for the background subtraction ($f=1.0$
and $f'=0.7$).

The surfaces covered by the regions C, NW and SE are 0.087468,
0.082725, and 0.060056 deg$^2$ respectively. The GLFs are plotted in
Fig.~\ref{fig:GLFregions} and the fit parameters are given in
Table~\ref{tab:GLFparam}.

In Region C, which corresponds to the cluster core, the importance of
the Gaussian component is strong, as it was for the overall cluster
(see Section~\ref{sec:GLFglobal}) and in the ``rectangle'' region. 
The faint end slope is somewhat flatter in region C than
in the overall cluster. Region C is therefore dominated by bright
galaxies, as region ``rectangle''.

In view of the shapes and poor qualities of the GLFs in regions NW and
SE, due to the low numbers of galaxies in these regions, no Gaussian
function needs to be included for $f=1.0$.  In fact a simple
power law would be sufficient to fit the GLFs, since in most cases
$M^*$ is not constrained. This implies that very bright galaxies are
concentrated in the inner zones of the cluster, as expected, and that
there are very few bright galaxies in the outer regions. The faint end
slopes of the GLFs in regions NW and SE are close to those found for
the overall image. We can also note the importance of the Galactic
extinction correction that somewhat modifies the parameters of the
GLFs, in particular in the SE zone, where the extinction has a larger
value (see Fig.~\ref{fig:mapexti}). For $f'=0.7$, the GLFs in
  regions NW and SE look much better, as expected, and the Gaussian
  component again appears necessary to account for the excess of
  bright galaxies.

\section{Discussion and conclusions}
\label{sec:concl}

\begin{figure}[h!]
 \begin{center}
\includegraphics[width=5.4cm, angle=-90]{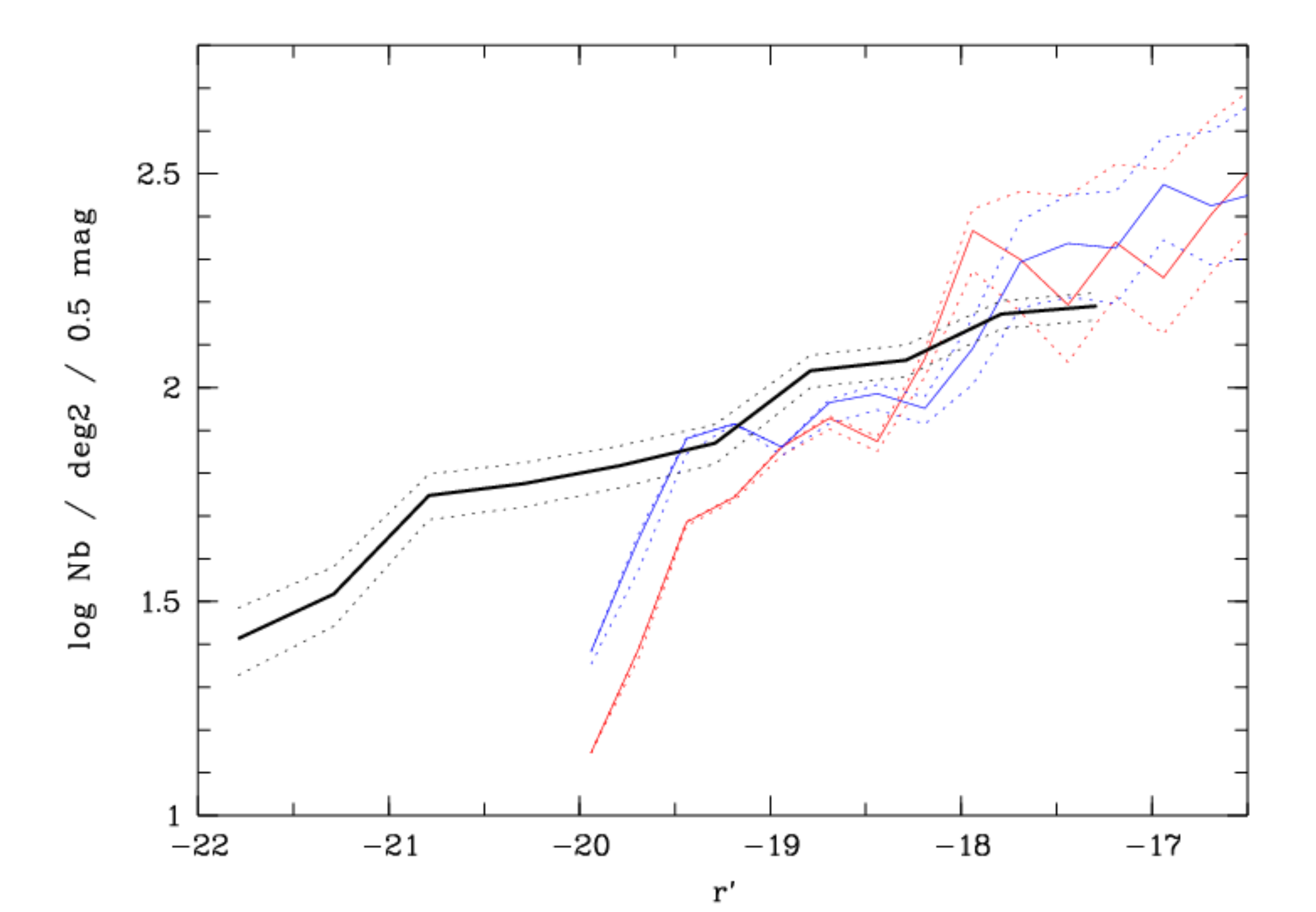}
\caption{Comparison of the GLFs of Ophiuchus (computed with
    variable extinction and $f=1.0$) and Coma in regions with the
  same physical sizes. The Ophiuchus GLF is in black, while the GLFs
  of the North and South zones of Coma are in red and blue,
  respectively.  The dotted lines show the $\pm 1\sigma$ errors on
  each curve.}
\label{fig:GLF_Oph_Coma}
\end{center}
\end{figure}

Due to its low Galactic latitude, the Ophiuchus cluster is difficult
to study at optical wavelengths. However, based on a photometric
catalogue of 2818 galaxies in a zone covering 1~deg$^2$ around
Ophiuchus, we have succeeded in obtaining the GLFs in the overall
image, in a region (the ``rectangle'' region) of the same physical
size as that considered for the Coma cluster by Adami et al. (2007),
as well as in three circular regions defined from density maps: a
central zone (C), and two regions northwest (NW) and southeast (SE) of
the centre. In all the regions, the GLFs show an excess of very bright
galaxies over a Schechter function, that is well fit by a
Gaussian. The faint end slope is quite flat in the very central region
C: between $-0.58$ and $-0.83$, while it is steeper in the outer
zones. However, we must note that the GLFs of zones NW and SE are
rather noisy, and that the choice of the correction for Galactic
extinction and of the factor by which to multiply the background
counts modifies the best fit parameters, so it is difficult to draw
conclusions for these zones.

Since Ophiuchus and Coma are two massive nearby clusters, it is quite
logical to compare their properties. The former is known to be quite
relaxed (see e.g. Paper~I) while extensive studies of the latter imply
that it is far from relaxed and still undergoing one or several
mergers. Therefore, the comparison of these two clusters can give us
informations on the effect the environment may have on their
properties.

As explained above, we have extracted the Ophiuchus GLF in a region
covering the same physical size (roughly $1.1\times 1.4$~Mpc$^2$) as
the region in which the Coma GLF was calculated by Adami et
al. (2007).  The GLFs of Ophiuchus (computed for $f=1.0$) and of
the North and South halves of Coma computed in regions of identical
physical size are shown in Fig.~\ref{fig:GLF_Oph_Coma}. We can see
from this figure that the GLFs are quite similar for galaxies fainter
than $M_{r'}\sim -19.5$, but that Ophiuchus shows a large excess over
Coma of galaxies brighter than $M_{r'}\sim -19.5$. The North region of
Coma is known to contain a large fraction of faint galaxies, while the
South region of Coma, which is in contact with a large scale filament
arriving from the South-West, contains more bright galaxies. Ophiuchus
contains many more bright galaxies than the South region of Coma, thus
requiring mergers of quite massive galaxies.

One explanation could be that in the core of Ophiuchus many galaxy
mergers take place. This agrees with the flatter faint end slope
of the GLF in region C, relatively to the two adjacent regions (NW and
SE), and relatively to the cluster at larger scale (in particular the
Overall region, see Table~\ref{tab:GLFparam}).

Mamon (1992) has shown that for a galaxy of mass $m$, the merging rate
in a cluster varies as $m^2/\sigma_v^3$, where $\sigma_v$ is the
cluster velocity dispersion. The respective velocity dispersions of
Ophiuchus and Coma are approximately 950~km~s$^{-1}$ (Paper~I) and
1200~km~s$^{-1}$ (Adami et al. 2009), so the ratio of the merging
rates in Ophiuchus and Coma for a galaxy of given mass $m$ is expected
to be $\sim (1200/950)^3=2.0$. At the absolute magnitude $M_r=-20$,
there are roughly 3 times as many galaxies in Ophiuchus than in Coma
(Fig.~\ref{fig:GLF_Oph_Coma}).  To obtain this factor of 3, the ratio
of the masses of bright galaxies in Ophiuchus and in Coma would
therefore need to obey the relation $(m_{Oph}/m_{Coma})^2 \sim 3/2$,
corresponding to bright galaxy masses in Ophiuchus $m_{Oph}$ larger
than the galaxy masses in Coma $m_{Coma}$ by about 20--25\%, values
which are plausible. This very simple order of magnitude calculation
shows that numerous galaxy mergers in the center of Ophiuchus can
indeed probably account for the very high number of galaxies brighter
than $\sim -19.5$.

Another explanation to the high number of mergers could be that
because Ophiuchus is embedded in a large scale system involving the
Great Attractor, this could favour galaxy--galaxy interactions and
mergers. Such mergers could involve bright field galaxies of
comparable masses and/or galaxies with small relative velocities,
favouring efficient energy exchange between these galaxies, and
forming even brighter merged objects.

We can note however that, though being a very rich and massive
cluster, Ophiuchus does not appear to be embedded in a very dense
network of galaxies, and its environment does not appear to be dense
(Wakamatsu et al. in preparation). On the other hand, the large scale
environment of Coma, which is also a very rich cluster, at a
comparable redshift, shows a much higher galaxy density.  In the
current paradigm where clusters are at the intersection of cosmic web
nodes, it is somewhat difficult to understand how two massive clusters
have built up to be of comparable mass and richness within such
different environments. The only explanation we see is that Ophiuchus
has built up long ago, as confirmed by its relaxed state (see
paper~I), while Coma is still in the process of forming, as
illustrated by the series of papers by Adami et al. (see e.g. Adami et
al. 2007). It would be interesting to test this hypothesis by
comparing the degree of relaxation and the large scale galaxy
distribution for a large sample of rich clusters.

\begin{acknowledgements}

  We thank the referee for her/his useful comments. We acknowledge
  enlightening discussions with Emmanuel Bertin.  We are extremely
  grateful to Andrea Biviano for giving us and adapting for us his
  program to fit the galaxy luminosity functions, for pointing out the
  Yasuda et al. (2001) paper, and for helpful discussions.  We thank
  Florian Sarron for his help in extracting the individual extinctions
  for all the galaxies and Vincent Caill\'e for help measuring some of
  the galaxy magnitudes. F.D. acknowledges long-term support from
  CNES.  J.M.O.M.M.M. was supported by the Brazilian agency CAPES
  (``Science without Borders'' program 88888.789740/2014-00).

This work is based on data obtained with MegaPrime/MegaCam (proposal
10AF02), a joint project of CFHT and CEA/DAPNIA, at the
Canada-France-Hawaii Telescope (CFHT) which is operated by the
National Research Council (NRC) of Canada, the Institute National des
Sciences de l'Univers of the Centre National de la Recherche
Scientifique of France, and the University of Hawaii.  This research
has made use of the NASA/IPAC Extragalactic Database (NED) which is
operated by the Jet Propulsion Laboratory, California Institute of
Technology, under contract with the National Aeronautics and Space
Administration.

\end{acknowledgements}

\appendix

\section{Typical errors on the magnitudes}

\begin{table}[h!]
\caption{ Typical errors on the measured magnitudes depending on the magnitude range.}
\begin{center}
\begin{tabular}{cc}
\hline
\hline
magnitude range & error on magnitude \\
\hline
       $r'< 20.0$  &      0.05\\
$20.0 < r' < 21.0$  &     0.10\\
$21.0 < r' < 22.0$  &     0.20\\
$22.0 < r' < 23.0$  &     0.5\\
$23.0 < r'$         &     0.8\\
\hline
\end{tabular}
\label{tab:errmag}
\end{center}
\end{table}

The typical errors on the measured magnitudes are given in Table~\ref{tab:errmag}
for various magnitude ranges.

\section{The first ten lines of the photometric catalogue}

The first ten lines of the photometric catalogue are shown in
Table~\ref{tab:cat}.

\small

\begin{table*}[h!]
  \caption{First ten lines of the photometric catalogue of Ophiuchus, which
    includes 2818 galaxies and is available in ViZieR at the
    CDS.
    The columns are: (1)~name,
    (2)~RA (2000.0), (3)~DEC (J2000.0), (4) and (5)~measured $r'$ band
    magnitude with no extinction correction, 
    (5) and (6)~magnitudes $r'_{ct}$ corrected for a
    constant value of 1.357~mag, and $r'_{var}$ corrected for the
    individual extinction correction, (7) and (8)~major axis $a$ and
    minor axis $b$ in arcseconds (the error bars on these two quantities are typically 
    0.1~arcsec, so they are not given for every galaxy),
    (9)~position angle $PA$ of the major axis, (10)~error on the position angle.
    We do not give individual errors on the observed magnitudes, but their typical values are
    given in Table~\ref{tab:errmag}.
  }
\begin{tabular}{rrrrrrrrrr}
\hline
\hline
                     &           &           &         &         &         &     &     &      &  \\
Name                 & RA        & DEC       & $r'$    &$r'_{ct}$ &$r'_{var}$& $a$ &$b$& $PA$ & $err_{PA}$ \\
                     & (J2000.0) & (J2000.0) &         &         &          &     &    &      &  \\
\hline
OPHJ171014.79-233403 & 257.56162 & -23.56767 & 15.655 & 14.298 & 14.558 & 75.2 & 18.0 & 168.3 & 0.1 \\
OPHJ171014.90-232009 & 257.56207 & -23.33589 & 17.686 & 16.329 & 16.473 & 15.9 & 11.9 & 122.2 & 3.0 \\
OPHJ171014.98-234752 & 257.56241 & -23.79794 & 20.455 & 19.098 & 19.103 &  6.4 &  3.6 &  31.1 & 0.1 \\
OPHJ171014.99-234735 & 257.56244 & -23.79315 & 21.253 & 19.896 & 19.888 &  4.4 &  4.2 &  60.4 & 0.6 \\
OPHJ171015.56-225946 & 257.56485 & -22.99636 & 21.844 & 20.487 & 20.720 &  4.6 &  2.3 &  93.6 & 0.1 \\
OPHJ171016.02-232204 & 257.56674 & -23.36787 & 20.407 & 19.050 & 19.300 &  4.9 &  3.5 & 159.2 & 0.1 \\
OPHJ171016.06-232500 & 257.56693 & -23.41680 & 22.074 & 20.717 & 21.049 &  3.7 &  3.1 &  25.3 & 0.2 \\
OPHJ171016.08-230323 & 257.56699 & -23.05655 & 21.401 & 20.045 & 20.266 &  5.1 &  3.4 &  53.0 & 0.8 \\
OPHJ171016.27-232231 & 257.56781 & -23.37537 & 20.211 & 18.854 & 19.150 &  4.9 &  3.5 & 162.1 & 0.6 \\
OPHJ171016.27-233547 & 257.56781 & -23.59645 & 20.141 & 18.784 & 18.992 &  5.5 &  3.8 & 110.1 & 0.1 \\
\hline
\end{tabular}
\label{tab:cat}
\end{table*}

\end{document}